\documentclass[trackchanges]{aastex701}

\usepackage{pifont} 
\newcommand{\cmark}{\ding{51}}%
\newcommand{\xmark}{\ding{55}}%

\usepackage{multirow}
\usepackage{booktabs}
\usepackage{amssymb}

\usepackage[version=4]{mhchem}

\begin{document}

\title{Powerful lightning on Venus constrained by atmospheric NO}

\author[orcid=0000-0002-2129-1340,gname=Tereza,sname=Constantinou]{Tereza Constantinou}
\affiliation{Institute of Astronomy, University of Cambridge}
\email[show]{tc496@cam.ac.uk}

\author[orcid=0000-0002-8713-1446,gname=Oliver,sname=Shorttle]{Oliver Shorttle}
\affiliation{Institute of Astronomy, University of Cambridge}
\affiliation{Department of Earth Sciences, University of Cambridge}
\email[show]{os258@cam.ac.uk}

\author[orcid=0000-0002-7180-081X,gname=Paul B.,sname=Rimmer]{Paul B. Rimmer}
\affiliation{Cavendish Astrophysics, University of Cambridge}
\email[show]{pbr27@cam.ac.uk}

\begin{abstract}

Signs of lightning on Venus have long been sought, including by space missions and ground-based telescopes searching for optical flashes, plasma waves, or radio signatures.  These efforts have yielded conflicting findings regarding the presence or absence of lightning in Venus's atmosphere. In this study we adopt an indirect approach to constrain the prevalence of lightning on Venus, using the chemical by-products it produces in Venus's atmosphere. Nitric oxide (NO) is a key tracer species of lightning, being exclusively generated by lightning in Venus's lower atmosphere. By calculating the present rate of atmospheric destruction of NO in Venus's atmosphere through photochemical-kinetic modelling, we constrain the lightning power required to sustain the estimated NO abundances on modern Venus. The reported NO constraints require lightning to generate at-least three times the power released on Earth; consistent with either a higher rate of strikes, or greater energy per strike, or a combination of both. Limited detections of optical flashes within the clouds could point to lightning striking deeper in the atmosphere and nearer the surface --- with the result that its optical flashes are obscured by the clouds --- driven by triboelectric charging during volcanic eruptions or wind interactions with surface sediments. Our findings underscore the importance for future missions of confirming lightning on Venus, either by verifying the below-cloud NO abundance, or by detecting another unambiguous lightning signature, to provide the first definitive evidence of lightning on a rocky planet other than Earth.

\end{abstract}



\section{Introduction} 
\linenumbers 

\label{sec:intro}

Non-equilibrium chemistry has long-been proposed as a plausible biosignature \citep{LOVELOCK1965, 1967Hitchcock, 1975Lovelock}. Yet, lightning, an abiotic process, can induce non-equilibrium chemical reactions \citep[e.g., atmospheric nitrogen fixation;][]{1980Hill}. Disentangling biological from abiotic signatures requires a systematic understanding of all non-biological processes capable of shaping atmospheric composition. Only by mapping this full abiotic landscape can we hope to identify true anomalies that may point to life \citep{constantinou2025comparative}.  Developing an understanding of non-equilibrium chemistry in atmospheres is therefore foundational to our investigation of rocky planets. In this study we investigate lightning, and how it can globally affect a planet's atmospheric chemistry.

To date, there is no definitive confirmation of lightning on another rocky planet --- either within or beyond our solar system. Lightning was first observed beyond Earth's atmosphere on Jupiter (a gas giant) in 1979 from photographic observations made by Voyager 1 \citep{BORUCKI1982492}. In the 1980s, Voyager 2 detected lightning on the ice giants; on Uranus \citep{zarka1986radio} through radio emission, and on Neptune \citep{burns1983saturn} by characterising observed electrostatic discharges. For a review of lightning on planets within our solar system, see \citet{2016Hodos}. As our neighbouring rocky planet, Venus provides a distinctive opportunity to examine the factors influencing lightning in a terrestrial environment.  

The presence and characteristics of lightning on Venus have been long debated (e.g., \citet{grebowsky1997evidence, borucki1982comparison, KRASNOPOLSKY200680, Lorenz2018} and references therein).  Several previous missions and observing campaigns have sought to constrain the presence and rate of lightning on Venus. These pursuits, each employing different observational methods, have resulted in contradictory conclusions.  This has left uncertainty over how universal lightning is as an atmospheric phenomenon on rocky planets and therefore how it may contribute to their atmospheric dis-equilibria.

\subsection{Lightning generating processes on rocky planets}
Lightning occurs following the build-up of electrical charge across regions of the atmosphere/surface of a planet, leading to the formation of ionised paths through an atmosphere that connect and discharge.  These discharge events and their associated release of electromagnetic radiation are the lightning bolts we are used to seeing on Earth. Lightning strikes at a global rate of 44$\pm$5\,s$^{-1}$ on Earth, with the majority of these discharges occurring within the clouds \citep{Christian2003}. Typical intra-cloud flashes occur between the upper-cloud positive charge and the primary negative charge regions in the middle to lower part of the cloud \citep{nssl_lightning_types}. Cloud-to-ground lightning occurs when electrical charge is exchanged between a cloud and the ground, with most strikes involving a downward transfer of negative charge (electrons), known as negative  lightning. Less frequently, positive cloud-to-ground lightning transfers positive charge downward and tends to be more energetic, comprising less than 5\% of all strikes \citep{NOAA_PositiveLightning}. 

While Venus may not have water-clouds to build up charge, it does have a global sulphuric acid cloud layer.  These sulphuric acid clouds could, in principle, build-up sufficient charge separation to trigger lightning. However, to do so, the Venusian clouds must encompass sulphuric acid in both solid and liquid form \citep{MCGOULDRICK2011934}. Therefore, any confirmed evidence of lightning on Venus would not only establish a key atmospheric energy source, but also provide critical constraints on the microphysical state and vertical structure of Venus’s clouds --- a long-standing and fundamental question in Venusian science.

Lightning is also produced on Earth during volcanic eruptions: in this case charge build-up occurs from friction between dry particles of ash and debris \citep{pahtz2010particle}, or the interaction between super-cooled cloud droplets, ice crystals, and falling graupel \citep[soft hail][]{Arason2011}. Discharges in volcanic ash clouds are facilitated by the elevated altitudes that expelled dust reaches, as the breakdown field (the minimum electric field strength needed to ionise a gas and create a conductive path for lightning discharge) is lower at high altitude, leading to more frequent, yet less energetic, discharges. 

Evidence for active volcanism on Venus includes an observed lava flow at a volcanic vent \citep{Herrick2023}, geologically young lava flows inferred from emissivity measurements \citep{smrekar2010recent, Filiberto}, and the requirement for ongoing volcanic degassing to sustain the global sulphuric acid (\ce{H2SO4}) cloud layer \citep{fegley1989estimation, my_venus_dry_paper}. The style of volcanism remains uncertain: pyroclastic deposits suggest that explosive eruptions capable of generating volcanic lightning may occur \citep{Ganesh2021}, though effusive volcanism may be more common \citep{smrekar2010recent, Herrick2023}. 

Historically, the absence of near-surface aerosols argued against explosive volcanic lightning \citep{borucki1982comparison, grebowsky1997evidence}. More recent work identifies a near-surface particulate layer (peaking at at 3.5--5\,km) with basaltic properties \citep{kulkarni2025search}, supported by earlier haze and Venera 13 observations \citep{anderson1969dust, seiff1995pioneer, mogul2023co2}. This reopens the possibility of localised volcanic discharges.

Aeolian processes may also contribute to lightning on Venus: winds transporting dust and sand can generate electrical charge via tribo-electrification \citep{lorenz2016surface, Lorenz2018}. Transported material could include basaltic dust  \citep{kulkarni2025search},  impact ejecta, or near-surface hazes \citep{woitke2025prediction}. Modelling predicts mineral sulphate hazes in the lower atmosphere (above 2–15\,km) \citep{woitke2025prediction}, closely matching haze layers seen by Pioneer Venus and Venera probes \citep{mogul2023co2}. These particles are also expected to carry high negative charge densities \citep[to prevent coagulation and agree with measured opacity data][]{woitke2025prediction}. Evidence of surface transport includes rearranged Venera 13 lander particles \citep{selivanov1982evolution} and wind-shaped features such as dunes and micro-dunes observed by Magellan \citep{greeley1992aeolian, weitz1994dunes, kreslavsky1998microdunes, lorenz2014dune}. Near-surface winds are non-uniform, with slope-driven diurnal winds strongest on steep low-latitude slopes \citep{DOBROVOLSKIS1993276}. Experiments and modelling suggest particles up to tens of microns can be lifted to several kms altitude, enabling widespread redistribution \citep{greeley1984windblown, greeley1990aeolian, sagan1975windblown}. Discharge current measurements near 1–2\,km indicate high particle charge densities \citep{Lorenz2018}, suggesting that eolian processes could produce localised electrical activity.

Venus’s high surface pressure presents a challenge for electrical discharges, with significantly higher breakdown electric field compared to Earth. The breakdown field --- the minimum electric field strength required to initiate a lightning discharge --- depends on atmospheric composition, pressure, temperature, and local charge density. Near Venus’s surface, the breakdown field is high at the compared to Earth \citep{RIOUSSET2020113506}.  As a result, while it may be harder to accumulate the necessary conditions for a discharge, any lightning that does occur is expected to be substantially more energetic than typical Earth lightning.

 \subsection{Searches for lightning on Venus}
A key technique to find lightning on Venus has been to look for optical flashes.  Optical flashes from lightning have a distinct peak at the excited atomic oxygen triplet near 777\,nm.  The 777\,nm feature was  searched for by several instruments, including the Venera 9 and 10 Orbiter Spectrometer \citep{krasnopolsky1983lightnings}, Pioneer Venus Star Tracker \citep{Borucki1991}, and Venus Express VIRTIS \citep{CARDESINMOINELO2016395}. These searches have not yet yielded clear detections (see summary of detection efforts in Table \ref{table:dec}).  Venera 9 recorded an optical flash with duration and optical energy comparable to Earth-like lightning, hinting to a lightning rate of 700\,s$^{-1}$ \citep{krasnopolsky1983lightnings}, though it is uncertain whether it was instead caused by instrumental anomalies or spacecraft debris \citep{Lorenz2018}. Two optical surveys, conducted with the Mt. Bigelow telescope \citep{HANSELL1995345} and the Akatsuki orbiter \citep[][article in review]{takahashi2021optical}, have captured several distinctive light flashes that could indicate lightning strikes. Such optical flashes, however, are not exclusive to lightning flashes as they could also be caused by impacting meteoroids \citep{Blaske}, potentially skewing detected lightning rates.

\begin{table*}
\centering
\begin{tabular}{llccccl}
\toprule
\multirow{2}{*}{Year} & \multirow{2}{*}{Mission or Observing Campaign} & \multicolumn{3}{c}{Detection Technique} & \multirow{2}{*}{Reference} \\
\cmidrule(lr){3-5}
 &  & Optical & Whistler & Radio & \\
\midrule
1975  & Venera 9 and 10 & \cmark / \xmark &       &        & \citenum{krasnopolsky1983lightnings}/\citenum{Lorenz2018} \\
1978  & Venera 11 and 12 &         &         & \textbf{?} & \citenum{krasnopolsky1979lightning}, \citenum{1980Ksanfomaliti} \\
1983  & VEGA             & \xmark  &         &        & \citenum{Sagdeev} \\
1990; 1978--1992 & Pioneer Venus Orbiter & \xmark & \cmark / \xmark & & \citenum{Borucki1991}; \citenum{scarf1983lightning}/\citenum{grebowsky1997evidence} \\
1991  & Galileo          & \xmark &         & \cmark / \xmark & \citenum{belton1991images}; \citenum{gurnett1991lightning}/\citenum{Lorenz2018} \\
1995  & Mt. Bigelow telescope$^a$ & \cmark / \xmark & & & \citenum{HANSELL1995345}/\citenum{Blaske} \\
1998, 1999 & Cassini     &         &         & \textbf{?} & \citenum{2001Gurnett} \\
2006--2014 & Venus Express & \xmark & \cmark / \xmark & & \citenum{CARDESINMOINELO2016395}; \citenum{RUSSELL20061344}/\citenum{George2023} \\
2011  & Calar Alto and Observatorio del Teide$^a$ & \xmark & & & \citenum{garcia13multi} \\
2021; 2019 & Parker Solar Probe & & \xmark & \textbf{?} & \citenum{Blaske}; \citenum{Pulupa2020} \\
2019--2022 & Akatsuki & \textbf{?} / \xmark & & & \citenum{Lorenz2019}/\citenum{Blaske} \\
\bottomrule
\end{tabular}
\caption{List of missions and observing campaigns that have sought for lightning on Venus using various detection techniques. A check-mark (\cmark) indicates a detection derived from lightning, a cross (\xmark) a non-detection or a detection exclusively from non-lightning processes, and a question mark (\textbf{?}) an inconclusive one. Symbols separated by `/` mean conflicting interpretations; `,` means different sources with same interpretation; `;` separates results from different techniques. $^a$ Ground-based.}
\label{table:dec}
\end{table*}

Lightning also generates very low frequency plasma waves during electrical discharge called whistler-waves, which propagate along magnetic field lines. The Venus Express magnetometer recorded magnetic bursts near Venus, initially interpreted as whistler-mode waves produced by lightning below the ionosphere \citep{RUSSELL20061344}. This interpretation suggested a flash rate several times higher than for lightning on Earth; up to 320 flashes per second \citep{2022Hart}. However, complicating this interpretation is recent work suggesting that whistler-waves were detected travelling towards Venus, and not away, so could not have been generated by lightning \citep{George2023}. 

Electrical discharges during lightning generate radio signals from the rapidly changing electric current \citep{Guo2022}. Despite several missions attempting to detect lightning-related radio signals during their Venus flybys, including Cassini \citep{2001Gurnett} and the Parker Solar Probe \citep{Pulupa2020}, such signals remained elusive. The Galileo flyby of Venus provided a few weak radio signals that were initially attributed to lightning \citep{gurnett1991lightning}.  However, these were later reinterpreted in favour of non-lightning explanations, including instrumental errors, plasma waves, and dust impacts \citep[Gurnett's reconsideration mentioned in][]{Lorenz2018}. In contrast to the above failures to detect lightning, the Venera 11–-14 landers observed near-continuous signals at low radio frequencies during their descents, resembling electrical activity associated with lightning on Earth \citep{1980Ksanfomaliti}. The Pioneer Venus Orbiter also detected electric fields and an optical anomaly thought to be linked to lightning flashes \citep{HuestisSlanger, 1979Taylor}.

 \subsection{Finding lightning through disequilibrium chemistry}
In contrast to the lightning detection methods discussed above, which were directly looking for the electromagnetic expression of lightning, lightning may also be inferred from its effect on the chemistry of an atmosphere. The atmospheric species NO, in particular, is a sensitive tracer of lightning processes, as lightning is thought to be the only known natural mechanism that could produce NO in the lower Venusian atmosphere \citep{KRASNOPOLSKY200680}.  This is because the dissociation of molecular nitrogen (\ce{N2}) --- which possesses a strong triple bond --- requires substantial energy input. While photodissociation can occur in the upper atmosphere under intense solar radiation, Venus’s lower atmosphere lacks sufficient thermal or photochemical energy to produce NO by conventional means. Thus, some other high-energy process must be invoked.

Early high-resolution spectroscopic observations revealed NO mixing ratios of 5.5\,ppb below 60\,km altitude, which was used to estimate a lightning flash rate of approximately 90\,s$^{-1}$ \citep{KRASNOPOLSKY200680}. This flash rate was determined by considering the NO flux necessary to support the pre-dissociation of NO and its interactions with N, adjusted for the production of NO and N resulting from cosmic ray ionization above the cloud layer. It is critical to note, however, that this remains the only reported NO detection, and it has yet to be independently confirmed. Building on this work, our photochemical-kinetic network additionally includes reactions between the N, H, C and S species, which are prevalent in the Venusian atmosphere. 

We also consider new observations that set upper limits on the abundance of NO above the clouds \citep{MAHIEUX2024115862}. In particular, \citet{MAHIEUX2024115862} report that the NO value above the cloud top is smaller by an order of magnitude than the 5.5\,ppb value estimated at 60\,km. 

In this paper we present a constraint on the lightning rate on Venus, by linking both observational constraints on NO and its lifetime in the atmosphere, to the lightning flash rate required to sustain it. Section \ref{sec:methods} describes the methodology, including the employed photochemical-kinetic model and the established connection between the atmospheric abundance of NO, and lightning activity. The relevant atmospheric chemistry and calculated NO flux are detailed in Section \ref{sec:results}, along with the derived lightning power on Venus. Discussions of the global lightning flash rate and energy per strike, particularly in light of past observations, are presented in Section \ref{sec:disussion}. Finally, our conclusions on the nature of Venusian lightning and potential implications for future missions are given in Section \ref{sec:conclusion}.

\section{Methods}\label{sec:methods}

\subsection{Estimating the NO production rate}\label{sec:methods_fluxes}
We model the behaviour of NO in the Venusian atmosphere (Figure \ref{fig:no_depth}) with a chemical-kinetic network that reproduces the modern atmosphere of Venus \citep{Rimmer2016, rimmer2021hydroxide}. We employ an atmospheric chemical network, STAND2022, which describes gas phase photo-chemical kinetics on Venus, and is used by ARGO, a 1D Lagrangian high energy photochemistry-diffusion model. STAND2022 includes over 6600 thermochemical and photochemical reactions involving 480 species composed of C-O-H-S-N-Cl and a handful of other elements (see \nameref{sec:data}). The chemical kinetics code has been validated to within an order of magnitude of most observations of atmospheric species abundances for Venus \citep{rimmer2021hydroxide}. The network robustly includes NO; the complete list of all reactions involving NO are listed across \citet{Rimmer2016} and \citet{rimmer2021hydroxide}. We discuss the particularly relevant reactions for this work in Section \ref{sec:results}.

Simulating the chemical journey of a gas parcel, ARGO tracks its ascent from the surface to the top of the atmosphere (115\,km) and back down. Within our atmospheric model, the bottom of the atmosphere is incorporated as fixed initial surface mixing ratios. These are chosen such that the model fits observational measurements of species abundances further up in Venus's atmosphere \citep{rimmer2021hydroxide}.  This is the approach taken for NO: its observationally constrained abundances within the Venusian atmosphere are matched by tuning its surface abundance in the model.  

As the gas parcel ascends, ARGO calculates the thermochemical equilibrium for gas-phase chemistry at each altitude step, as a function of the altitude's pressure and temperature \citep{rimmer2021hydroxide}. Upon reaching the upper atmosphere, ARGO incorporates the incident stellar spectrum. Descending through the atmospheric layers, it solves for both thermochemical equilibrium and the photochemical reactions driven by the stellar flux at each height step. The time spent evolving the chemistry at each height-step depends on the timescale of vertical transport, as dictated by the eddy diffusion profile \citep[from][]{2007Krasnopolsky, Krasnopolsky2012}. The choice of eddy diffusion levels within the cloud layer is investigated in Section \ref{sec:results}.

Measurements of Venus's atmospheric chemistry over the twentieth and twenty-first centuries establish minimum timescales for the atmospheric evolution of the planet, and are ultimately consistent (within error) with Venus having a broadly stable atmosphere over this period \citep{2023NatCoHelbert, my_venus_dry_paper}. Assuming the atmosphere of Venus is in steady-state, any atmospheric species being destroyed through atmospheric processes (i.e., photochemistry and thermochemistry) would require a replenishing source through processes not considered in the photochemical-kinetic network. These processes include exogenic delivery, surface-atmosphere interactions, or lightning. For NO, lightning stands out as the sole source beyond the chemical reactions already accounted for in the kinetic network.  

It is important to note that there is only one claimed detection of NO to date \citep{KRASNOPOLSKY200680}, and thus no direct proof of a true steady state in the mixing ratio of this key lightning tracer. However, even if the \citet{KRASNOPOLSKY200680} detection captured NO at a transient maximum immediately after production, some process must still have occurred to generate it. Therefore, regardless of whether NO is steady-state or transient, any net destruction of NO the model predicts implies at least a production flux (if not `restorative' flux to maintain steady-state) equal in magnitude.

The rate of change of an atmospheric species $X$ at each height $z$ in the atmosphere, is determined with a set of time-dependent, coupled, non-linear differential equations:
\begin{equation}\label{eq:partial}
\frac{\partial n_X}{\partial t} = P_X - L_Xn_X - \frac{\partial \Phi_{\mathrm{diff,}X}}{\partial z},
\end{equation}
at height $z$ (cm) from the surface, where $n_X$ (cm$^{-3}$) is the number density of a species $X$, $P_{X}$ (cm$^{-3}$ s$^{-1}$) is the rate of production of species $X$, $L_{X}$ (s$^{-1}$) is the rate constant for destruction of species $X$, and $\Phi_{\mathrm{diff,}X}$ (cm$^{-2}$\,s$^{-1}$) is the vertical flux of species as a function of eddy- and molecular-diffusion processes. $P_{X}$ and $L_{X}$ are determined from the rate constants of the different reactions within STAND, and the relevant species abundances \citep{Rimmer2016, rimmer2021hydroxide}.

The net flux of $X$ in the atmosphere, $\Phi_{atmo, {X}}$ (cm$^{-2}$ s$^{-1}$), is determined by summing its chemical production and destruction within an air parcel throughout its circulation cycle in ARGO. This calculation integrates the aforementioned photochemical, thermochemical, and diffusive processes within the atmosphere
\begin{equation} \label{eq:full_flux_eqn} 
 \Phi_{\mathrm{atmo,} X} = \int_{0}^{115\,km} \left[P_{X}(z,t)- L_{X}(z,t)n_{X}(z,t) - \frac{\partial \Phi_{\mathrm{diff,}X}}{{\partial z}}\right] dz, \\
\end{equation}
where 115\,km is the maximum height considered in ARGO. 

As ARGO is a Lagrangian model and does not track fluxes,  $\Phi_{\mathrm{atmo,} X}$ is determined by multiplying the change in the species abundance $\Delta n_{X}$ (cm$^{-3}$) across the history of the parcel with the velocity of the modelled gas parcel \citep{my_venus_dry_paper}; this serves as a proxy for the time-integrated flux when considering all photochemical, thermochemical and diffusive processes. $\Delta n_{\ce{X}}$ is assessed from the initial ascent of the gas parcel cycle until it descends to the surface and reaches convergence. The velocity of the gas parcel at the surface is derived from the eddy diffusion $K_{ZZ}$ (cm$^{2}$ s$^{-1}$) at the base of the atmosphere, and the atmospheric scale height $H_{sc}$ (cm) through $\frac{2K_{ZZ}}{H_{sc}}$, such that
\begin{equation} 
\label{eq:endflux_eqn} 
 \Phi_{\mathrm{atmo,} X} = \frac{2 \Delta n_{X} K_{ZZ}}{H_{sc} }. \\
\end{equation}
From this, the e-folding time of atmospheric species is determined from their expected lifetime in the atmosphere, defined as the species column density divided by $\Phi_{atmo,X}$. 

The dominant reactions that produce or destroy X are determined by comparing the rates $R_{chem}$ (cm$^{-3}$ s$^{-1}$) of individual reactions. At each altitude step, the reaction rate is calculated at the last time-instance that the parcel spends at each height $z$,  
\begin{equation}\label{eq:rate} 
R_{chem} =  k(z) \prod_{i = 1}^{N} n_{X_{i}}\\
\end{equation}
where $k(z)$ (cm$^{3N-3}$ s$^{-1}$) is the reaction rate constant extracted from STAND right before the parcel moves, N is the total number of reactants, $\prod_{i = 1}^{N} n_{i}$ (cm$^{-3N}$) is the product of the number density of all N reactants. The rates capture a ``snapshot" of the parcel's chemistry at the last step before the parcel moves, effectively identifying the dominant reactions.

\subsection{Estimating the power of lightning}\label{sec:estim_power}
On Earth, most lightning occurs within clouds, with about a third of lightning occurrences being cloud-to-ground discharges. The energy released per lightning flash varies based on the flash type and the stage of the storm, ranging from $2\,\times\,10^{8}$ to $7\,\times\,10^{9}$\,J per flash \citep{Maggio2008}.  Intra-cloud flashes, on average, yield an estimated energy of $1.6\,\times\,10^{9}$\,J, whereas cloud-to-ground lightning flashes release about half of that energy \citep{Maggio2008}.  We adopt the more energetic Earth-derived flash energy of $1.6\,\times\,10^{9}$\,J as the average energy dissipated by Venusian lightning flashes, similar to the $1\,\times\,10^{9}$\,J previously adopted for Venus \citep{KRASNOPOLSKY200680}. This is to act as a somewhat conservative estimate of the lightning flash rate on Venus, as our model only allows us to constrain the total power of lightning (thus assuming larger average energy would equate to a lower total rate). If lightning on Venus is dominantly occurring at ground level where the breakdown field is high, the energy liberated per flash could be higher still.

Lightning in Venus's atmosphere can produce nitric oxide (NO) with an experimentally measured efficiency of $(3.7\pm0.7)\times10^{15}$ molecules\,J$^{-1}$ \citep{1982Levine}. The process is initiated by energetic electrons that ionise and dissociate atmospheric species, primarily through the dissociation of molecular nitrogen, forming transient species such as \ce{N2+}, \ce{N+}, and \ce{N-}. However, the dominant nitrogen chemistry is effectively captured by the dissociation of molecular nitrogen,
\begin{align}
\ce{N2} + \ce{e-} &\rightarrow 2\ce{N} + \ce{e-},\label{eq:diss_N2}
\end{align}
though this may proceed through short-lived ionised or excited intermediates.

Atomic nitrogen readily reacts with atomic oxygen, sourced via photodissociation or lightning-induced dissociation of \ce{CO2}, the predominant molecule in Venus's atmosphere, to form NO via three-body recombination
\begin{equation}
\ce{N} + \ce{O} + \ce{M} \rightarrow \ce{NO} + \ce{M}.
\end{equation}
\citet{Krasnopolsky2012} also considered the radiative attachment reaction as a means to produce or restore \ce{NO},
\begin{equation}
\ce{N} + \ce{O}  \rightarrow \ce{NO} + hv,
\end{equation} 
emitting in the $\gamma$- and $\delta$-bands of NO at 190–300\,nm \citep{krasnopolsky1983lightnings}. The radiative attachment has a rate constant of $1.9 \times 10^{-17} {\rm cm^3 \, s^{-1}} \; \big(T/300 \, {\rm K}\big)^{-0.5}$ \citep{dalgarno1992radiative}, which corresponds to a restoration time of \ce{NO} on the order of millions of years, which is far too slow to restore any measurable amount of \ce{NO} above the clouds (See also Figure \ref{fig:test_react}).

Lightning can also dissociate \ce{N2} to excited state \ce{N}($^2$\ce{D}) atoms, which can subsequently react with \ce{CO2}, the most abundant species in Venus's atmosphere, to form \ce{NO}
\begin{align}
&\ce{N2} + \ce{e-}  \rightarrow  2\ce{N}(^2\ce{D}) + \ce{e-},\\
&\ce{N}(^2\ce{D}) + \ce{CO2}  \rightarrow \ce{NO} + \ce{CO}.
\end{align}
See \citet{DELITSKY2015184} and \citet{Qu2023} for a thorough overview of the complete list of electrochemical reactions that may occur. 

On Earth, abundant \ce{O2} supports a self-sustaining NO production chain
\begin{align}
\ce{N} + \ce{O2} &\rightarrow \ce{NO} + \ce{O},\\
\ce{O} + \ce{N2} &\rightarrow \ce{NO} + \ce{N}.
\end{align}
This forms a self-propagating cycle: each \ce{N} atom reacts with \ce{O2} to produce \ce{NO} and an \ce{O} atom, which then reacts with \ce{N2} to produce more \ce{NO} and regenerate \ce{N}, allowing the chain to continue.

In contrast, Venus's CO$_2$-rich atmosphere lacks free O$_2$, preventing such runaway chemistry. Although atomic nitrogen can react with CO$_2$
\begin{equation}
\ce{N} + \ce{CO2} \rightarrow \ce{NO} + \ce{CO},
\end{equation}
there is no analogous reaction to regenerate nitrogen atoms, limiting the build-up of NO.

To estimate the total power dissipated by lightning on Venus, we scale the NO production rate per unit energy $S_{\ce{NO}}$ by the required atmospheric NO flux $\Phi_{\ce{NO}}$ (cm$^{-2}$\,s$^{-1}$). The global power  released by lightning on Venus is then 
\[
P = \frac{\Phi_{\ce{NO}} \times 4\pi R_{\rm Venus}^2}{S_{\ce{NO}}} \,\, \text{[J\,s$^{-1}$]},
\]
(using $R_{\text{Venus}}$ in\,cm). From this, the average energy per flash ($E$) and global flash rate ($\nu$) follow: $P = E \times \nu$.

\section{Results}\label{sec:results}
\subsection{Photochemical-kinetics}
\begin{figure*}[htbp]\centering
	\includegraphics[width=\columnwidth]{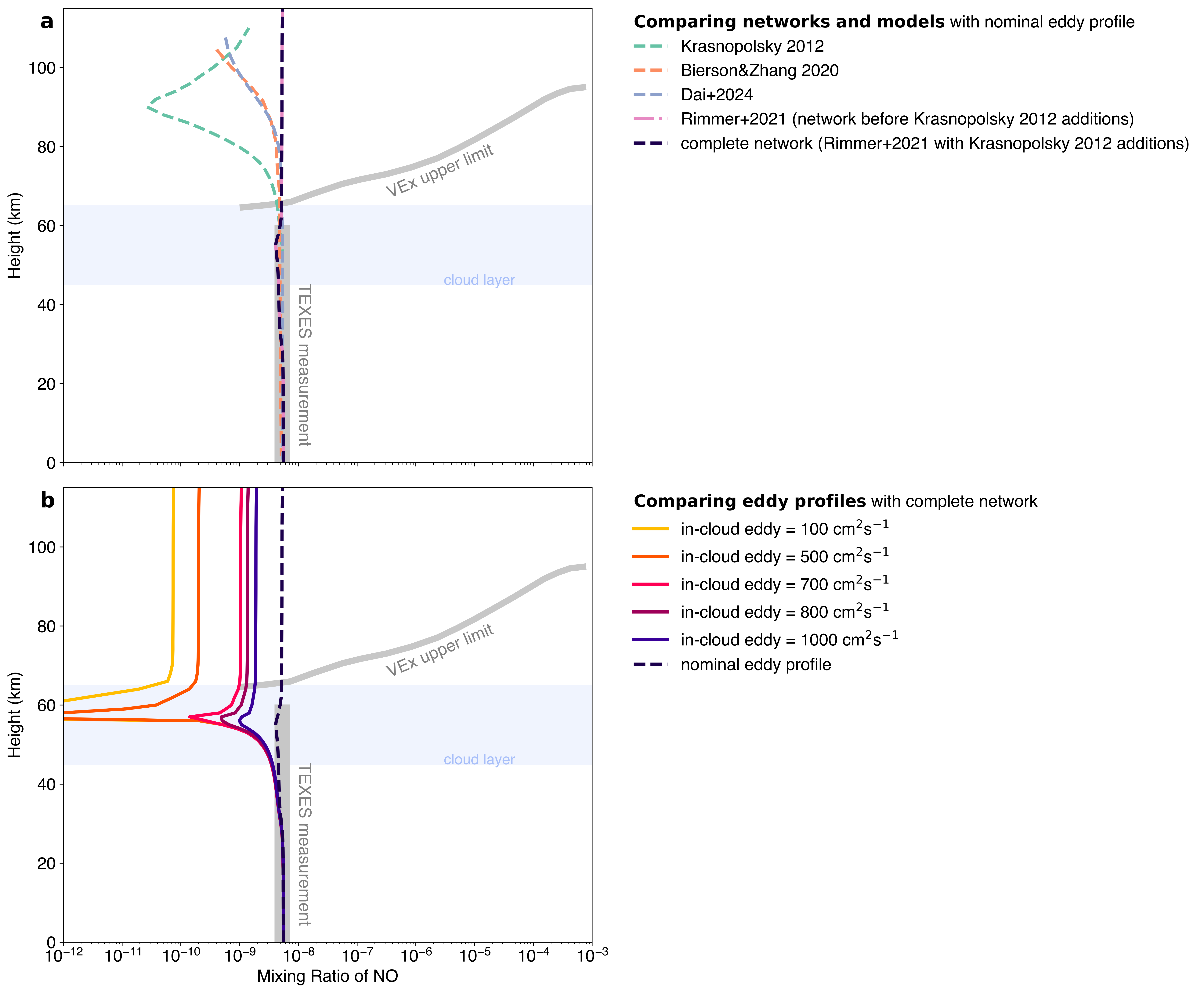}
 \caption{Modelled NO mixing ratios in Venus's atmosphere compared with observational constraints. \textbf{a,} Comparison of atmospheric chemistry models. Our complete model (black solid line) integrates the \citet{rimmer2021hydroxide} network with additional reactions from \citet{Krasnopolsky2012}. Other models shown include those from \citet{Krasnopolsky2012} (turquoise dashed), \citet{bierson2020chemical} (blue dashed), \citet{dai2024investigation} (orange dashed), and the original \citet{rimmer2021hydroxide} network without additions (pink dashed, primarily hidden under black dashed line). Grey shaded regions represent NO measurements from the TEXES spectrograph at NASA’s Infrared Telescope Facility \citep{KRASNOPOLSKY200680} and upper limits derived from non-detections by Venus Express (VEx) \citep{MAHIEUX2024115862}. All models exceed the upper limits in the cloud top region ($\sim$60--70~km), underscoring a persistent discrepancy between model predictions and observations. 
\textbf{b,} Sensitivity of the complete network model to variations in the in-cloud eddy diffusion coefficient (100--1000~cm$^{2}$~s$^{-1}$). The nominal profile (black dashed) over-predicts NO abundances above the cloud layer, while reduced in-cloud mixing (coloured lines) lowers NO to within the VEx upper limits. eddy profiles used are plotted in Supplementary Figure \ref{fig:eddy_profs}.}
 \label{fig:no_depth}
\end{figure*}

The chemical-kinetic network we use has been validated to within an order of magnitude of most observations of atmospheric species abundances for Venus \citep{rimmer2021hydroxide}. However, recent upper limits on NO derived from Venus Express non-detections \citep{MAHIEUX2024115862} revealed an over-prediction of NO above the clouds, not only in our model, but systematically across published Venus models (Figure \ref{fig:no_depth}, upper panel).

To investigate this discrepancy, we explored whether adjusting the in-cloud eddy diffusion coefficient ($K_{\text{zz}}$) could bring the modelled NO profile into agreement with observations.  Vertical mixing in Venus’s cloud region is poorly constrained due to the absence of reliable direct measurements \citep{Dai2025}. One-dimensional atmospheric models, like the one used here (Section \ref{sec:methods_fluxes}),  parametrise complex three-dimensional dynamics with $K_{\text{zz}}$. This coefficient is typically constrained by vertical motions and abundance gradients of trace species in the presence of sufficient observational constraints. We use a similar approach that varies $K_{\text{zz}}$ to match \ce{NO} observational constraints. 

Most one-dimensional Venusian atmospheric chemistry-transport models assume relatively low values of $K_{\text{zz}}$ in the cloud region, on the order of $10^4\,\mathrm{cm^2\,s^{-1}}$ \citep{Krasnopolsky2012, zhang2012sulfur, shao2020revisiting, bierson2020chemical, dai2024investigation}. \citet{bierson2020chemical} used values as low as $10^3\,\mathrm{cm^2\,s^{-1}}$ to reproduce SO$_2$ profiles, despite independent evidence for strong convective activity and low static stability in this region \citep{imamura2017initial, ando2020thermal, oschlisniok2021sulphuric}, which would suggest higher mixing efficiencies. This discrepancy illustrates how $K_{\text{zz}}$ values tuned to match one species may not capture the full dynamical complexity of the three-dimensional cloud environment \citep{lefevre2022impact, lefevre2024impact, Dai2025}.

Alternative modelling approaches point to significantly higher eddy diffusion coefficients. Using VEx observations of H$_2$SO$_4$ and assuming condensation--diffusion equilibrium, \citet{dai2023determination} derived in-cloud $K_{\text{zz}}$ values up to $10^8\,\mathrm{cm^2\,s^{-1}}$, with evidence of strong latitudinal variation. Turbulence-resolving simulations coupled to chemical models by \citet{lefevre2022impact, lefevre2024impact} derived values of $K_{\text{zz}}$, ranging from $10^6$ to $10^8\,\mathrm{cm^2\,s^{-1}}$; also substantially higher than those adopted in most one-dimensional models.

Reducing in-cloud $K_{\mathrm{zz}}$ increases the chemical residence time of the gas parcel within the clouds, allowing for more time for reactions to take place, and enhancing NO depletion via in-cloud reactions before upward transport. We find that with our photochemical-kinetic model, the modelled NO is sufficiently depleted below the Venus Express upper limit for in-cloud $K_{\mathrm{zz}} < 700$\,cm$^2$\,s$^{-1}$ (Figure~\ref{fig:no_depth}, lower panel). This provides a self-consistent way to reconcile known NO chemistry with available observations.

That said, the required low $K_{\mathrm{zz}}$ should not be interpreted as a literal constraint on in-cloud mixing. Turbulence-resolving models predict much higher $K_{\mathrm{zz}}$ values \citep{lefevre2022impact, lefevre2024impact}, indicating that while our low value provides an effective parametrisation of unresolved processes, it does not reflect the true mixing efficiency in Venus’s clouds. More generally, the exercise illustrates the uncertainty in compressing 3D transport into a 1D framework: $K_{\mathrm{zz}}$ values tuned to a single species may successfully reproduce that species’ behaviour, but they cannot be regarded as definitive constraints on the wider cloud environment.

We define the best-fit eddy profile as the maximum in-cloud $K_{\mathrm{zz}}$ for which the model satisfies both the observed lower-cloud NO abundance \citep[5.5~$\pm$1.5\,ppb, ][]{KRASNOPOLSKY200680} and the upper-limit constraint above the clouds, achieved through depletion within the cloud layer. With this adjustment, our chemical-kinetic model remains consistent with the observed behaviour of nitrogen species --- and other species, plotted against observations in Supplementary Figure~\ref{fig:comp_all_obs} --- throughout the atmospheric column. This enables us to extract the dominant reaction pathways controlling nitrogen speciation on Venus.

\subsection{NO flux from vertical mixing ratio gradient}
The prevalent form of nitrogen on Venus is molecular nitrogen (\ce{N2}), constituting $\sim$3\% of the Venusian atmosphere, making it the second most abundant gas. \ce{N2} is long-term stable, exhibiting an e-folding time of $4.0\,\times\,10^{7}$ years (Section \ref{sec:methods}). Its dominant chemical reactions (Supplementary Figure \ref{fig:reactions_N2}) lead to the creation of temporary radicals in the deep atmosphere ($\lesssim$20\,km) such as \ce{NCO}, \ce{N2O}, \ce{H2N}, through reactions 
\begin{align}
& \ce{N2 + CO2 -> NO + NCO} &
k &= 8.05 \times 10^{-11}\,\mathrm{cm^3\,s^{-1}}
\left(\frac{580\,\mathrm{K}}{300\,\mathrm{K}}\right)^{-1.98}
\exp\!\left(\frac{7.1 \times 10^{4}\,\mathrm{K}}{580\,\mathrm{K}}\right)
\text{ (ref.~\citenum{1992tsang})},\\[6pt]
& \ce{N2 + CO2 -> CO + N2O} &
k &= 5.30 \times 10^{-13}\,\mathrm{cm^3\,s^{-1}}
\exp\!\left(\frac{6.4 \times 10^{4}\,\mathrm{K}}{580\,\mathrm{K}}\right)
\text{ (ref.~\citenum{tsang1991chemical})},\\[6pt]
& \ce{N2 + H2O -> NO + H2N} &
k &= 1.28 \times 10^{-10}\,\mathrm{cm^3\,s^{-1}}
\left(\frac{580\,\mathrm{K}}{300\,\mathrm{K}}\right)^{-2.65}
\exp\!\left(\frac{1.04 \times 10^{5}\,\mathrm{K}}{580\,\mathrm{K}}\right)
\text{ (ref.~\citenum{miller1999modeling})}.
\end{align}

where the rate constants are given for 580\,K at an altitude of 20\,km. These radicals have a short-lived existence in the atmosphere, however, as they readily revert to \ce{N2} with similar reaction rates. 

While the production of NO by lightning acts as a sink of atmospheric \ce{N2}, the chemical-kinetics of the atmosphere allow the conversion of NO back into \ce{N2} through the combined reaction pathway and net reaction
\begin{align}
&\ce{H2S + hv -> HS + H}, \label{eq:photodissociation} \\
&\ce{HS + HS -> H2S + S}, \\
&\frac{1}{8} \left(\ce{8S -> S8}\right), \label{eq:elemental-sulfur} \\
&\ce{H + CO + M -> HCO + M}, \label{eq:hco_formation} \\
&\ce{HCO + NO -> CO + HNO}, \label{eq:hco_no_reaction} \\
&\ce{HNO + CO -> CO2 + NH}, \label{eq:hno_oxidation} \\
&\ce{NH + NO -> OH + N2}, \label{eq:nh_no_reaction} \\
&\ce{CO + OH -> CO2 + H}, \label{eq:co_oxidation} \\
&\ce{2H -> H2}, \label{eq:h2-generation} \\
&\text{Net reaction:} \quad 
\ce{H2S + 2NO + 2CO -> 2CO2 + N2 + H2} + \frac{1}{8}\ce{S8}. \label{eq:net_reaction}
\end{align}
with corresponding rates for the bimolecular and termolecular reactions of,
\begin{align*}
&\text{Reaction}~\ref{eq:elemental-sulfur} 
&&k = 1.50 \times 10^{-11} \,\mathrm{cm^3\,s^{-1}}
&\text{ref.~\citenum{schofield1973evaluated}},\\
&\text{Reaction}~\ref{eq:hco_formation} 
&&k = 1.96 \times 10^{-13} \,\mathrm{cm^3\,s^{-1}} 
&\text{ref.~\citenum{wagner1987addition}},\\
&\text{Reaction}~\ref{eq:hco_no_reaction} 
&&k = 1.18 \times 10^{-11} \,\mathrm{cm^3\,s^{-1}} 
\exp\!\left(-\tfrac{1.37 \times 10^{3}\,\mathrm{K}}{T}\right)
&\text{ref.~\citenum{dammeier2007wide}},\\
&\text{Reaction}~\ref{eq:nh_no_reaction} 
&&k = 5.86 \times 10^{-12} \,\mathrm{cm^3\,s^{-1}} 
\left(\tfrac{T}{300\,\mathrm{K}}\right)^{-0.5}
&\text{ref.~\citenum{BOZZELLI1994965}},\\
&\text{Reaction}~\ref{eq:co_oxidation} 
&&k = 1.43 \times 10^{-13} \,\mathrm{cm^3\,s^{-1}} 
\left(\tfrac{T}{300\,\mathrm{K}}\right)^{1.55}
&\text{ref.~\citenum{tsang1986chemical}},
\end{align*}

This net reaction network depletes NO in the cloud layer, by effectively catalysing the conversion of NO and CO into \ce{N2} and \ce{CO2}, with \ce{H2S} acting as a reductant and itself being atomised. Ultimately, the sulfur ends up as \ce{S8}, as represented by reaction~\ref{eq:elemental-sulfur}, though the specific pathway can be complex. Atomic hydrogen can end up in many different species, but one is in the formation of molecular hydrogen, as illustrated by reaction~\ref{eq:h2-generation}. The rate-limiting step is Equation~\ref{eq:nh_no_reaction} (reaction rates that make up the net depletion mechanism plotted in Supplementary Figure \ref{fig:rate_limigint}), with rate constant $k = 5.86 \times 10^{-12} \, \mathrm{cm^3\,s^{-1}} \left(\frac{T}{300\,\mathrm{K}}\right)^{-0.5}$ \citep{BOZZELLI1994965}. Reactions \ref{eq:hno_oxidation}, \ref{eq:nh_no_reaction} and \ref{eq:co_oxidation} have not been previously considered in atmospheric NO flux calculations on Venus \citep{KRASNOPOLSKY200680}.

\begin{figure}[htbp]
	\includegraphics[width=\columnwidth]{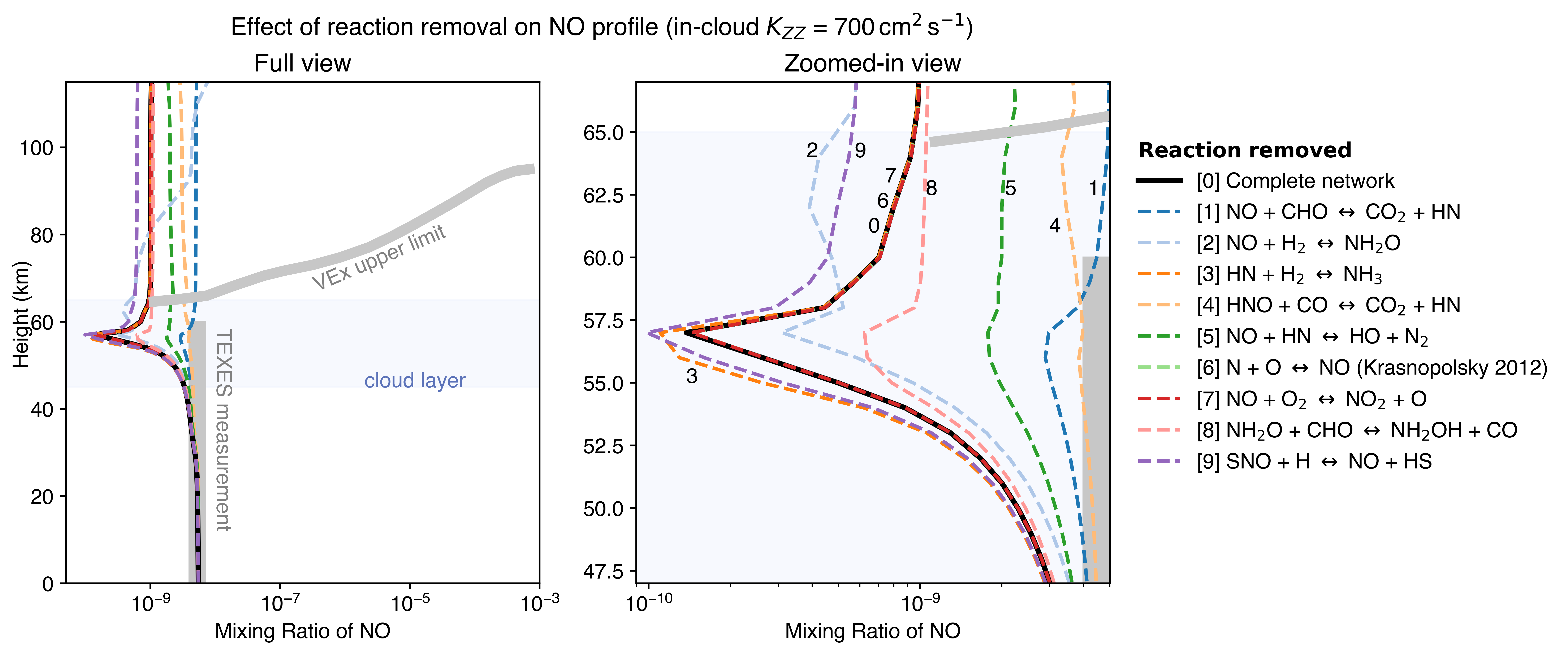}
\caption{
Sensitivity of modelled NO mixing ratios to individual reactions in the chemical network. Profiles show the NO mixing ratio as a function of altitude when specific reactions (and their reverse reactions) are removed from the network to assess their impact. \textbf{Left,} are the complete profiles, and \textbf{Right,} profiles are zoomed-in the cloud area where NO is depleted.  The solid black line shows the baseline result with the full chemical network included.  Grey shaded regions indicate observational constraints: NO measurements from the TEXES spectrograph \citep{KRASNOPOLSKY200680}, and upper limits from non-detections by VEx \citep{MAHIEUX2024115862}. Recombination reaction \ce{N + O <-> NO} was not removed, but instead its reaction rate was replaced with that of  \citet{Krasnopolsky2012} for comparison ( replacement rate coefficient is listed in Table~\ref{table:kr12reactions}).
}
 \label{fig:test_react}
\end{figure}

To confirm that this pathway is the depletion mechanism, we compare the NO profile from our complete network, to networks with each reactions from the net reaction removed in Figure \ref{fig:test_react}. The lack of, or greatly decreased, in-cloud depletion for cases where a reaction from the pathway is removed evidences that this net reaction is the primary depletion mechanism. This destruction pathway for NO is in agreement with results by \citet{spacek2023production}, where they explain that the abundances of HCO and HCHO both depend critically on the amount of NO in the atmosphere --- HCO and NO abundances are co-dependent via Equation \ref{eq:hco_no_reaction}.

\begin{figure}[htbp]
	\includegraphics[width=\columnwidth]{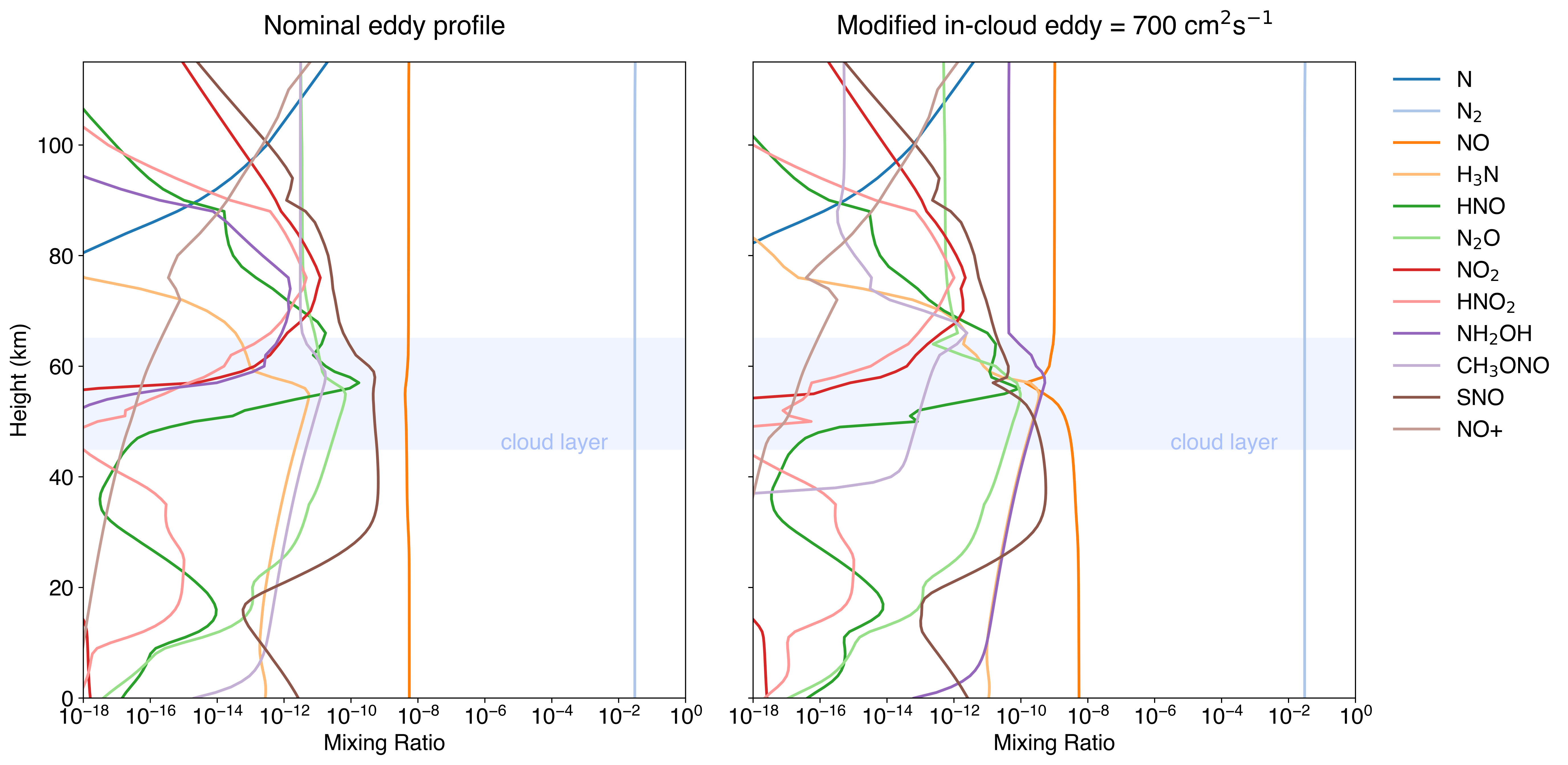}
 \caption{Vertical profiles of the most abundant nitrogen-containing species as a function of altitude in our model. \textbf{Left,} Nominal eddy diffusion profile used in \citet{rimmer2021hydroxide}, which overestimates NO abundances relative to observational constraints. \textbf{Right,} Modified eddy profile with reduced in-cloud diffusion (700~cm$^{2}$~s$^{-1}$), consistent with NO upper limits from Venus Express (VEx) \citep{MAHIEUX2024115862}.}
 \label{fig:n_spec}
\end{figure}

Other than \ce{N2}, some NO is also temporarily converted to \ce{NH2OH} in the clouds (peak in \ce{NH2OH} mixing ratio concurrent with \ce{NO} dip in cloud layer in Figure \ref{fig:n_spec}). This is because of the increased hydrogen abundance we have in our model \citep[from theorised hydroxide salts][]{rimmer2021hydroxide}, via
\begin{align}
&\ce{NO + H2 -> NH2O} 
&&k = 1.00 \times 10^{13} \,\mathrm{cm^3\,s^{-1}}\,
\exp\!\left(-\tfrac{5.5 \times 10^3\,\mathrm{K}}{350}\right) 
&\text{ref.~\citenum{2006Li}}, \label{eq:h2no} \\
&\ce{NH2O + HCO -> NH2OH + CO} 
&&k = 1.54 \times 10^{-5} \,\mathrm{cm^3\,s^{-1}}\,
\left(\tfrac{T}{300\,\mathrm{K}}\right)^{-2.15}
&\text{ref.~\citenum{xu2004computational}}, \label{eq:nh2oh_formation}
\end{align}
where the rate constant for reaction \ref{eq:h2no} is evaluated in the clouds ($\sim$50\,km) at 350\,K. Though most of \ce{NH2O} reverts back to \ce{H2} and \ce{NO}, restoring NO above the clouds, a fraction reacts with \ce{HCO} to form \ce{NH2OH}. If there is less H available within the clouds, less NO is expected to be restored above the clouds. This pathway, however, is not the key depletion mechanism for \ce{NO}: the in-cloud depletion of NO is only minimally reduced by the removal of Equation \ref{eq:h2no} from the network, shown in Figure \ref{fig:test_react}. Other reactions that are operating at fast rates in the atmosphere (Supplementary Figure \ref{fig:no_react}) are also removed for verification that they are not part of the primary depletion mechanism. In our network (though likely not in reality), \ce{NH2OH} is not destroyed at the cold temperatures of the cloud layer. As a result, it accumulates until it reaches warmer regions near the surface (see Figure \ref{fig:n_spec}), where it decomposes thermally into 
\begin{align}
&\ce{NH2OH -> NH2 + OH} 
&&k = 6.77 \times 10^{-10} \,\mathrm{cm^3\,s^{-1}} \,  
\left(\frac{T}{300\,\mathrm{K}}\right)^{-3.23} 
\exp\left(-\frac{2.6 \times 10^4\,\mathrm{K}}{350}\right) 
&\text{ref.~\citenum{1999Herron}}, \label{eq:NH2OH} \\ 
&\text{and then } \ce{NH2 + HCl -> NH3 + Cl} 
&&k = 1.08 \times 10^{-11} \,\mathrm{cm^3\,s^{-1}} \,  
\exp\left(\frac{2.6 \times 10^4\,\mathrm{K}}{350}\right) 
&\text{ref.~\citenum{gao2006thermochemistry}}, \label{eq:NH2hcl}
\end{align}
where both reactions are evaluated at an altitude of $\sim$5\,km at 700\,K. 

Interestingly, some ammonia (\ce{NH3}) is also formed as a minor bi-product of \ce{NO}'s depletion pathway. The HNO made via Equation \ref{eq:hco_no_reaction} can also subsequently react with CO leading to 
\begin{align}
&\ce{HNO + CO -> CO2 + NH} 
&&k = 3.32 \times 10^{-12} \, \mathrm{cm^3\,s^{-1}} \, 
\exp\left(-\frac{6.19 \times 10^3\,\mathrm{K}}{T}\right) 
&\text{ref.~\citenum{rohrig1994reactions}}, \\ 
&\text{and } \ce{NH + H2 -> NH3} 
&&k = 3.62 \times 10^{10} \, \mathrm{cm^3\,s^{-1}} \, 
\left(\frac{T}{300\,\mathrm{K}}\right)^{-1.00} \, 
\exp\left(-\frac{5.5 \times 10^3\,\mathrm{K}}{T}\right) 
&\text{ref.~\citenum{hanson2018survey}}, \label{eq:nhh2nh3}  
\end{align}
where reaction \ref{eq:nhh2nh3} is evaluated in the clouds ($\sim$50\,km) at 350\,K.

Considering all the production and destruction reactions, the \textit{net} atmospheric flux of NO in the atmosphere is a destruction flux of 2.0\,$\times$\,10$^8$cm$^{-2}$\,s$^{-1}$. Assuming the atmosphere is in steady-state, this necessitates an equal resupplying flux for NO to sustain a fixed number density in the atmosphere. Nitric oxide's atmospheric e-folding time of $3.2\,\times\,10^{3}$ years (Section \ref{sec:methods_fluxes}) is short on geological timescales, thus supporting the need for continual replenishment through some non-photochemical process (i.e., a process outside the atmospheric model used). Notably, both thermal- and photo-dissociation of \ce{N2} (the primary nitrogen reservoir) are negligible: thermal dissociation is ineffective throughout the atmosphere, and photodissociation only occurs above 100\,km. Given the detection of significant NO at a much lower altitude \citep{KRASNOPOLSKY200680}, this likely has an alternate source. Lightning currently stands as the sole known candidate for such a non-photochemical process, though further investigation is necessary to elucidate the specific mechanism and reconcile these observations.

Even if NO is not assumed to be continually supplied, there remains no other proposed source that could have supplied it so recently in Venus's past. NO is therefore strong evidence for the presence of lightning on Venus. 

We can demonstrate the model-independence of the NO destruction result by comparing the flux estimates from running the full photochemical kinetic calculation with a simple gradient-based approach. To estimate an NO flux independent of the chemical network, we apply a simple gradient-based calculation using the observational constraints. A $\sim$5\,ppb change in NO across the 60-64\,km altitude range implies a required minimum flux of $2.7 \times 10^8$\,cm$^{-2}$ s$^{-1}$, assuming eddy diffusion is the dominant transport mechanism, and drawing from the nominal eddy profile ($1.5 \times 10^4$\,cm$^{2}$ s$^{-1}$ average eddy coefficient in that height range; Figure~\ref{fig:eddy_profs}). This flux is derived using the relation
\begin{align}
    \Phi_{\text{NO}}=  \frac{ \Delta n_{\text{NO}}  H_{\text{sc}}}{\tau_{K_{ZZ}}},
\end{align}
where $\tau_{K_{ZZ}}$ is the eddy diffusion timescale across that altitude step.  Accounting for the broad range of proposed eddy coefficients ($K_{zz}$ = 700\,cm$^{2}$ s$^{-1}$ constrained here, up to $5 \times 10^8$\,cm$^{2}$ s$^{-1}$ from \citet{dai2023determination}), this network-independent flux could range from $7.9 \times 10^7$ -- $5.6 \times 10^{13}$\,cm$^{-2}$ s$^{-1}$.
This observationally-necessitated range is in agreement with the model-derived fluxes described here, supporting the need for an active NO source.

\subsection{Other lightning-produced species}
Electrical discharge in an experimentally-simulated Venusian atmosphere also produced \ce{CO} and \ce{O_n} through the dissociation of \ce{CO2} \citep{DELITSKY2015184}. Our atmospheric model, however, does not reflect this as the photochemical-kinetics produces a surplus of CO \citep{my_venus_dry_paper}, necessitating some form of deposition as opposed to production through lightning. The model also predicts \ce{O_n} is in a quasi-steady state, not requiring a source from lightning. Unlike NO, solely produced by lightning, CO and \ce{O_n} have more diverse chemical cycles. \ce{CO} could have additional fluxes into or out of the atmosphere through volcanism and weathering reactions, and \ce{O_n} likely experiences significant atmospheric escape \citep{2014Gillmann}. Ultimately the atmospheric CO and \ce{O_n} atmospheric fluxes are involved in further processes not accounted for by the model, and as a result we do not consider these species further.

\section{Discussion}\label{sec:disussion}

\subsection{Meteoroid shock chemistry as a source of NO}
Before making the connection to lightning, it is important to consider other high-energy processes that could contribute to NO formation. Meteoroids enter Venus's atmosphere, and create luminous superheated shock layers \citep[hence their invocation as a cause of the Akatsuki flash by][]{Blaske}, which could also drive disequilibrium shock synthesis of nitrogen oxides. Laboratory experiments using laser-induced dielectric breakdown provide analogues for such shock chemistry \citep{ferus2019main, ferus2023simulating}, with yields of $\sim$1.3\,$\times$\,10$^{16}$ NO molecules\,J$^{-1}$ of deposited energy in Venus-like gas mixtures \citep{mvondo2001production, heays2022nitrogen}. \citep[This is higher than the $\sim$3.7\,$\times$\,10$^{15}$\,molecules\,J$^{-1}$ reported for lightning-discharge experiments][]{1982Levine}.  

Taking an upper-limit meteoroid mass flux of 3\,kg\,s$^{-1}$ \citep{peucker1996accretion} and adopting a mean entry velocity of $\sim$25.6 km,s$^{-1}$ \citep{le2011nonuniform}, the total energy input would be $\sim$2\,$\times$\,10$^{9}$\,W. Applied to the laboratory yield, this corresponds to $\sim$2\,$\times$\,10$^{25}$ molecules\,s$^{-1}$ of NO. Deposited above the clouds, this introduced NO would be destroyed (chemical reaction rate of $\sim$1 molecules\,cm$^{-3}$\,s$^{-1}$), preventing any build-up; thus the observed above-cloud upper limit is consistent with meteoroid chemistry.

Even under the most favourable assumption --- that all meteoroid energy deposition occurs within or below the clouds --- this rate still falls at least an order of magnitude short of the $\sim$9\,$\times$\,10$^{26}$ molecules\,s$^{-1}$ required to sustain the reported detection \citep{KRASNOPOLSKY200680}. So while meteoroid shocks may sporadically contribute to the Venusian NO budget, they cannot plausibly account for the below-cloud NO abundance \citep{KRASNOPOLSKY200680}.

A further possibility is episodic production from a past large airburst or impact. Assuming a conservative diffusion-limited lifetime for NO of  $\sim$1\,$\times$\,10$^{10}\,$s, then a $\sim$ $\sim$1\,$\times$\,10$^{21}\,$J airburst that happened $\sim$300 years ago could account for the NO observed \citep[as also calculated for phosphine by][]{bains2021phosphine}. However, the very low frequency of such large impacts \citep{quintana2016frequency} makes this explanation statistically unlikely. 

\subsection{Photochemical estimates of the rate of lightning on Venus}
Lightning emerges as the most reliable candidate for restoring the NO that is continually destroyed by atmospheric chemistry. To counteract this destruction flux and sustain the reported NO abundance, and accounting for the experimentally-derived yield of NO production per Joule under Venusian conditions \citep[Section \ref{sec:estim_power};][]{1982Levine}, a global power of 2.4\,$\times$\,10$^{11}$\,J\,s$^{-1}$ is required to be released from lightning on Venus. 

On Earth, lighting generates an average global power of 7.0\,$\times$\,10$^{10}$\,J\,s$^{-1}$, $\sim$3 times less than we infer for Venus from NO. This difference could either be accounted for by an increased global rate of lightning on Venus, or more energetic lightning strikes. 

Assuming an Earth-like average strike energy of 1.6\,$\times$\,10$^{9}$\,J, the best fit of the modelled NO profile to observations in Venus's atmosphere requires a global average rate of $\sim$150 lightning strikes per second. However, determining the optimal solution for energy per strike and lightning rate to explain global lightning power on Venus remains an ongoing challenge, with a range of potential solutions depicted in Figure \ref{fig:energy_rate}. The best-fit model characterises the lightning power required to sustain the observed NO levels in Venus' atmosphere; spanning a range of lightning energies and rates with lower limits imposed by Earth's lightning: with a strike energy of 1.6\,$\times$\,10$^{9}$\,J, and rate of 44$\pm$5\,s$^{-1}$ \citep{Christian2003}. The analysis extends to encompass a broader range of solutions, incorporating the fluxes derived from the NO depletion described by observational constraints, and the range of possible in-cloud $K_{zz}$ values (Section \ref{sec:results}). These cover a range of possible solutions, necessitating lightning of power 9.8\,$\times$\,10$^{10}$--7.0\,$\times$\,10$^{16}$ \,J\,s$^{-1}$, so even more frequent and/or energetic strikes. 

The atmospheric profile of NO thus suggests lightning on Venus could be at-least $\sim$3 times more frequent than on Earth \citep[contingent on the below-cloud detection of][]{KRASNOPOLSKY200680}. A previous study also modelling the behaviour of NO in the Venusian atmosphere, similarly deduced a global flash rate of $\sim$90\,s$^{-1}$ \citep{KRASNOPOLSKY200680}, based on a smaller calculated NO flux of 7\,$\times$\,10$^{7}$\,cm$^{-2}$\,s$^{-1}$. However, their atmospheric flux of NO was derived considering only atmospheric N-chemistry, and the NO profile was only validated using the TEXES below-cloud measured abundance of NO.  An updated version of this model which includes N-C-H-S-O chemistry \citep[Figure \ref{fig:no_depth};][]{Krasnopolsky2012} estimated the NO flux to be 1\,$\times$\,10$^{10}$\,cm$^{-2}$\,s$^{-1}$ \citep{Krasnopolsky2012}, equivalent to a lightning wattage of $\sim$1\,$\times$\,10$^{13}$\,J\,s$^{-1}$ ($\sim$40 times greater than our estimate in this study). Our model, which incorporates additional N-C-H-S-O reactions \textit{and} is in agreement with the new upper limit on above-cloud NO abundance \citep{MAHIEUX2024115862}, places NO flux measurements between these two values.

\subsection{The lack of direct lightning detection on Venus}
It is perhaps surprising that NO abundances on Venus can imply such high lightning rates, given that missions seeking to detect optical flashes have not been successful \citep{CARDESINMOINELO2016395}. Cloud scattering reduces the optical energy reaching higher-altitude probes: flashes near the cloud base spread into 300\,km-wide bursts at the cloud tops, with only $\sim40\%$ of the original optical energy transmitted \citep{williams1983optical}. While this partial attenuation alone would not prevent detection by orbiters or Earth-based telescopes, near-surface flashes transmit a mere 0.01\% of their visible optical energy to space \citep{WILLIAMS1982166}, making cloud-to-ground or volcanic lightning effectively undetectable at present. This could explain the contrasting results of non-detection of optical flashes from space, while whistler-mode waves suggest higher occurrence rates of lightning \citep{2022Hart}. 

Detections of whistler mode waves have been used to infer a global lightning flash rate of up to 320\,s$^{-1}$ on Venus \citep{2022Hart}, almost twice the rate of flashes we require to sustain atmospheric NO abundances (assuming Earth-like strike energies). Additional non-lightning sources are therefore required to explain the full set of observations \citep{gurnett1991lightning}. However, as noted previously, Parker Solar Probe observations indicate that perhaps none of the observed whistler waves on Venus are from lightning \citep{George2023}. 

Ground-based observations from Earth of optical flashes on Venus set a minimum optical energy limit of $1.2\,\times\,10^{9}$\,J for each flash \citep{KRASNOPOLSKY200680}. However, there is no definitive way to confirm whether these flashes originated from lightning or meteor impacts. Assuming the observed flashes are indeed attributable to lightning, scaling the minimum optical energy to the total energy released per flash implies that lightning events on Venus could be 60--250 times more energetic  \citep{hansell1995optical, borucki1996spectral} than those on Earth \citep{Maggio2008}. This surpassed limits set by our model, but could be in agreement with the broader possible range allowed by other high in-cloud $K_{zz}$ estimates and the observed NO depletion (Figure \ref{fig:energy_rate}). However, we can only confidently draw conclusions about the extraordinary energy potential of Venusian lightning by first attributing these flashes to lightning by ruling out other possibilities. 

\subsection{Lightning on Venus and Earth}
\begin{figure*}[htbp]\centering
	\includegraphics[width=0.7\textwidth]{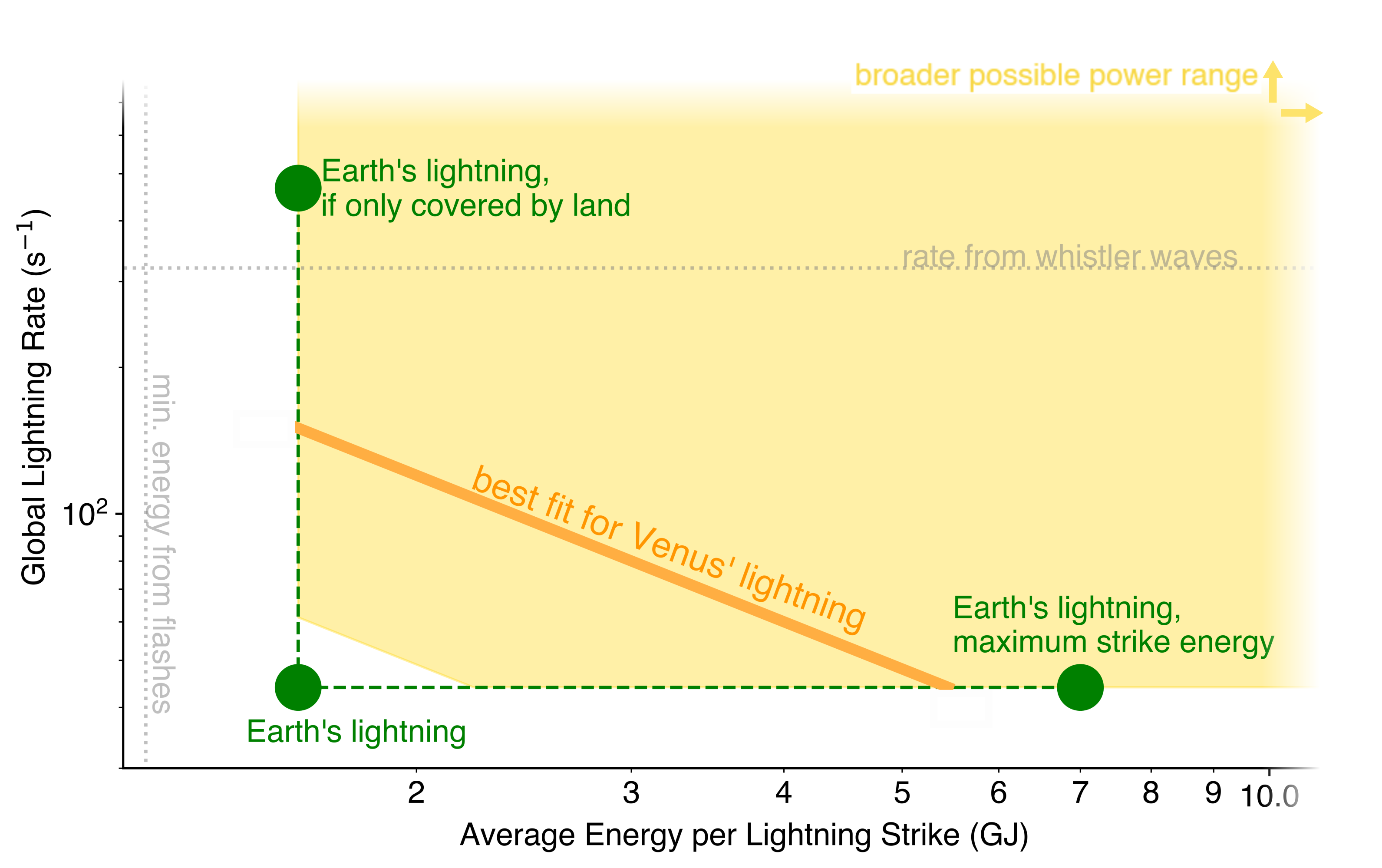}
 \caption{Lightning energy and rate values consistent with the required lightning power for Venus. The best-fit curve (best agreement with observational constraints on atmospheric abundance of NO, orange line) is plotted as derived from our atmospheric photochemical-kinetic model, with in-cloud $K_{zz}$=700\,cm$^{2}$\,s$^{-1}$.  The broader possible power range, as determined from the minimum depletion flux from  observational constraints and alternative in-cloud $K_{zz}$ values, extends to higher energies and rates than the plotted range (orange area filled-in; Section \ref{sec:results}). For reference, also plotted is the Earth's average and  energy per lightning strike \citep{Maggio2008} with the observed lightning rate \citep{Christian2003}, and the equivalent rate if the entire surface was only made up of land, without any oceans \citep{2016Hodos} (lower and upper green circles on the left, respectively), and the average rate with the maximum energy per lightning strike  \citep{Maggio2008} (right-most green circle). Also indicated are the global lightning flash rate of up to 320\,s$^{-1}$ inferred from whistler mode waves \citep[horizontal grey dotted line][]{2022Hart}, and minimum optical energy limit of $1.2\,\times\,10^{9}$\,J from on ground-based observations of optical flashes on Venus \citep[vertical grey dotted line][]{KRASNOPOLSKY200680} (neither are confirmed to be linked to lightning). }
 \label{fig:energy_rate}
\end{figure*}

The position of Venus's lightning parameters in Figure~\ref{fig:energy_rate} --- falling within the range bounded by Earth’s observed lightning energies and the theoretical upper limit for an ocean-free, all-land Earth --- suggests that lightning on Venus could be broadly Earth-like in character. This alignment implies that our terrestrial understanding of lightning may be extendable to exoplanets, even those with extreme atmospheric conditions like Venus. On Earth today, lightning contributes a small but important fraction of fixed nitrogen to the atmosphere \citep[e.g.,][]{Schumann2007}. Laboratory simulations of an anoxic, \ce{CO2}- and \ce{N2}-rich Archean Earth --- comparable to Venus’s present atmosphere --- have shown that lightning can produce substantial quantities of fixed nitrogen, with lightning producing a \ce{NO} flux of 6\,$\times$\,10$^{8}$\,cm$^{-2}$\,s$^{-1}$ before it is re-stabilised as \ce{N2} \citep{2019Delgado}. On Venus, our model suggests that lightning similarly induces nitrogen fixation, at a lower flux (2.0\,$\times$\,10$^{8}$\,cm$^{-2}$\,s$^{-1}$; Section \ref{sec:results}).  The less effective conversion of electrical energy into \ce{NO} in Venus’s atmosphere \citep{1982Levine} is what necessitates more powerful lightning given the observed NO mixing ratio.  Although \ce{NO} ultimately reverts to \ce{N2} (Equation \ref{eq:net_reaction}; Figure~\ref{fig:n_spec}), sustained lightning input maintains the observed NO levels, positioning lightning as a persistent source of chemical disequilibrium --- potentially a more dominant driver than on Earth.

On Earth, approximately 70\% of lightning occurs within clouds, with the remainder consisting of cloud-to-ground flashes \citep{Christian2003}.  By contrast, the lack of definitive optical flash detections on Venus from orbit suggests that if lightning occurs, it may take place deeper in the planet's atmosphere --- potentially below the main cloud deck --- where it would be obscured from view. Such discharges could manifest as lower-intra-cloud, cloud-to-surface, or near-surface lightning.

Unless a charge separation mechanism can operate efficiently within the sulphuric acid cloud --- such as freezing and melting of droplets \citep{MCGOULDRICK2011934} --- cloud-based lightning may be infrequent. The inference that discharges occur in the lower cloud layers or even deeper in the atmosphere stands in tension with proposed charging mechanisms that rely on the presence of sulphuric acid ice, which only forms from the middle of the cloud-deck and higher (above $\sim$56\,km) where temperatures fall below the freezing point of \ce{H2SO4}. The acidic nature of the cloud particles, with their high proton content (low pH), could inhibit the spatial separation of charges needed to initiate lightning.

Classical cloud-to-ground lightning is also difficult to envisage occurring on Venus, as the primary cloud deck is separated from the surface by $\sim$45\,km \citep[][contrast with Earth's $<$18\,km cloud deck]{NOAA2023clouds}. If lightning occurs in the lower atmosphere, it may be driven by non-meteorological mechanisms, such as volcanic or eolian (wind and dust-related) activity. 

Volcanic lightning, while observed on Earth, remains poorly quantified and is typically omitted from global lightning surveys. This makes it difficult to extrapolate to Venus, especially given the uncertain rate and style of its volcanism. Eolian lightning is even less understood. It is not observed on Earth, likely due to high humidity near the surface, which increases conductivity and suppresses triboelectric charging \citep{2020Harper}. Venus's much drier near-surface environment, however, may enable such charging processes, though this remains speculative without direct observation.

Another contributing factor to an increased lightning rate on Venus could be its completely land-covered surface. On Earth, cloud-to-ground lightning is significantly more frequent over land than over oceans. In the absence of oceans, Venus may not experience the same suppression of lightning over large regions, potentially resulting in globally higher lightning rates \citep{2016Hodos}. This is illustrated in Figure~\ref{fig:energy_rate}, which compares lightning energy and rate combinations consistent with the NO abundance observed in the Venusian atmosphere.

An alternative explanation to the powerful lightning on Venus could involve lightning strikes that are less frequent, but more energetic, as anticipated by the large breakdown field near the surface \citep{RIOUSSET2020113506}. To reproduce lightning rates similar to Earth's, lightning strikes on Venus would have to be up to $\sim$3 times more energetic \citep[though could be even more energetic;][]{RIOUSSET2020113506}. This is in agreement with our expectation of lightning on Venus being more prevalent in the form of eolian of volcanic near the surface where the breakdown field is larger, compared to further up in the clouds with intra-cloud or cloud-to-surface lightning. 

The mechanism of charge separation, responsible for generating lightning on Earth, could operate differently on Venus. This could allow for lightning as we do not know it, and could allow for more energetic lightning strikes to explain the large power necessitated by Venus's atmospheric NO inventory. Other than a larger breakdown field with increased charge separation before a strike, more branched electrical discharge paths could also lead to more energetic lightning strikes. However, experimental studies for lightning under Venusian atmospheric conditions, both within sulphuric acid clouds and at near-surface conditions, are necessary to test this hypothesis.

\subsection{Implications for future Venus missions}
If lightning is present on Venus, it could pose a hazard for future missions. Confirmation could come from independent detection of below-cloud NO \citep{KRASNOPOLSKY200680}, other chemical tracers, new optical/electromagnetic observations (see Table \ref{table:dec}), or, ideally, a combination of these methods. Only once lightning is unambiguously verified can its risks and implications for mission design be meaningfully assessed.

Confirmed lightning would pose little threat to short-lived or cloud-top missions, but long-duration platforms within or below the lower cloud deck, as well as surface probes, would face higher risks. Early Venus probes were not reported to be struck by lightning, but their rapid failure to heat and pressure means they provided limited exposure time \citep{Siddiqi2018}. Future aerial platforms or landers, particularly those targeted to geologically active regions, could face non-negligible strike probabilities over mission lifetimes.

In that case, mission designs could incorporate shielding and surge resilience, as was implemented for the Huygens probe \citep{mccarthy1997lightning, lebreton2002huygens}. Probes like DAVINCI \citep{Garvin_2022} and proposed landers such as Venera-D \citep{2018Zasova} would need similar protections to prevent damage to structural components or electronics. Strategically targeting the near-surface environment with lightning-hardened instruments would both safeguard hardware and maximize the chance of directly detecting lightning—resolving one of the longest-standing open questions in Venus atmospheric science.

\section{Conclusion}\label{sec:conclusion}
While various lines of observational evidence, such as optical flashes and whistler-mode waves, have been presented as potential indications of lightning on Venus, the question of whether lightning indeed occurs on Venus remains unresolved. Nitric oxide is generated exclusively by lightning discharges in the lower atmosphere of Venus.  This makes NO's presence a key indicator of lightning processes and provides the route we explore to quantifying the prevalence of lightning on Venus. 

Our calculations show that the NO flux necessary to sustain NO at the reported below-cloud abundance in the Venusian atmosphere could be supplied with lightning. This lightning would release a third of the global lightning power released on Earth. Given the distinct environmental conditions influencing lightning rates across Earth and Venus, both more frequent or more energetic events may also be consistent with the NO production rate. Limited optical flash observations within Venus's clouds hint at lightning produced lower in the atmosphere, attenuated before reaching space: either as low-cloud-to-ground discharges, or near the surface driven by volcanic or eolian tribo-electrification.

Crucially, our findings underscore the urgent need for future missions or observing campaigns to prioritise the confirmation and characterisation of NO in the deep atmosphere of Venus. Verifying the \citet{KRASNOPOLSKY200680} measurement would not only validate the powerful conclusions that can be drawn from atmospheric chemistry, but would also provide one of the strongest pieces of evidence to date for the existence and intensity of lightning on our neighbouring planet.

Our conclusions are readily verifiable through upcoming Venus missions focused on lightning detection, like ISRO's Venus Orbiter Mission \citep{ISRO2025LIVE}. We highlight that the most accurate lightning rate measurements would be made below the clouds, especially by incorporating simultaneous optical flash and radio measurements \citep[as initially suggested by][]{Blaske} to avoid ambiguity from meteoroids, which do not produce a radio signal. To constrain the lightning source, measurements within the clouds, and near active volcanic sites and large, steep slopes at low latitudes are encouraged.

\section*{Data Availability}\label{sec:data}
The data used to produce and validate the atmospheric model can be found across: \citet{Rimmer2016, RIMMER2019124, rimmer2021hydroxide, hobbs2021sulfur}. The data needed to evaluate the findings has been deposited in the Harvard online database: in \citet{DVN_19HF6Q_2024} and under accession code \url{https://doi.org/10.7910/DVN/R2GGAS}. 

\section*{Code Availability}
The methods underlying the code and the chemical network are comprehensively documented across papers: \citet{Rimmer2016, rimmer2021hydroxide, RIMMER2019124, my_venus_dry_paper}. The ARGO software with the atmospheric flux calculation extension may be made available upon request.

\begin{acknowledgments}
 T.C. thanks the Science and Technology Facilities Council (STFC) for the PhD studentship (grant reference ST/X508299/1). T.C. thanks Alex Archibald for useful discussions on atmospheric chemistry, and Jaidev Kenth and Skyla White for help with photochemical cross sections and the chemical network. 
\end{acknowledgments}

\begin{contribution}
TC implemented the simulations, analysed the output data, and wrote the manuscript. 
OS and PBR helped conceived the study and edited the paper.

\end{contribution}

%

\software{Matplotlib \citep{Hunter2007},  
          Pandas \citep{reback2020pandas}
          }


\appendix


\begin{table}[h!]
\centering
\renewcommand{\arraystretch}{1.5}
\begin{tabular}{llc}
\hline
\textbf{Reaction} & \textbf{Rate Coefficient} & \textbf{Units} \\
\hline
\ce{NO + HO2 -> NO2 + OH} & $k = 3.50 \times 10^{-12}  e^{-250/T}$ & cm$^3$\,s$^{-1}$ \\
\ce{NO2 + O -> NO + O2} & $k = 5.10 \times 10^{-12} e^{-210/T}$ & cm$^3$\,s$^{-1}$ \\
\ce{ClNO + Cl -> Cl2 + NO} & $k = 5.80 \times 10^{-11}  e^{-100/T}$ & cm$^3$\,s$^{-1}$ \\
\ce{ClNO + O -> NO + ClO} & $k = 8.30 \times 10^{-12}  e^{1520/T}$ & cm$^3$\,s$^{-1}$ \\
\ce{NO + ClO -> NO2 + Cl} & $k = 6.20 \times 10^{-12}  e^{-295/T}$ & cm$^3$\,s$^{-1}$ \\
\ce{NO + ClO2 -> ClNO + O2} & $k = 4.50 \times 10^{-11}  $& cm$^3$\,s$^{-1}$ \\
\ce{NO2 + SCl -> NO + OSCl} & $k = 2.30 \times 10^{-11}  $& cm$^3$\,s$^{-1}$ \\
\ce{ClCO + NO2 -> CO2 + NO + Cl} & $k = 6.00 \times 10^{-13}e^{-600/T}$& cm$^3$\,s$^{-1}$ \\
\ce{ClCO + ClNO -> COCl2 + NO} & $k = 8.00 \times 10^{-11} e^{573/T}$& cm$^3$\,s$^{-1}$ \\
\ce{N + O -> NO} & $k = 1.90 \times 10^{-17} \left(\frac{T}{300\,\mathrm{K}}\right)^{0.50} $& cm$^3$\,s$^{-1}$ \\
\ce{NO + Cl + M -> ClNO + M} & $k = 2.00 \times 10^{-31} \left(\frac{T}{300\,\mathrm{K}}\right)^{-1.80} $& cm$^6$\,s$^{-1}$ \\
\ce{NO2 + O + M -> NO3 + M} & $k = 2.43 \times 10^{-31} \left(\frac{T}{300\,\mathrm{K}}\right)^{-1.80} $& cm$^6$\,s$^{-1}$ \\
\hline
\end{tabular}
\caption{Reactions from \citet{Krasnopolsky2012} (see references therein) added to our chemical network to form our ``complete network''. All reactions are reversed in the network. Reaction \ce{N + O -> NO} is a radiative association reaction that replaces the three-body association from \citet{rimmer2021hydroxide}, which used $k = 7.87 \times 10^{-14} \left(\frac{T}{300\,\mathrm{K}}\right)^{-1.00}$ (cm$^6$\,s$^{-1}$).
 }
\label{table:kr12reactions}
\end{table}

\begin{figure}[htbp]
	\includegraphics[width=0.9\columnwidth]{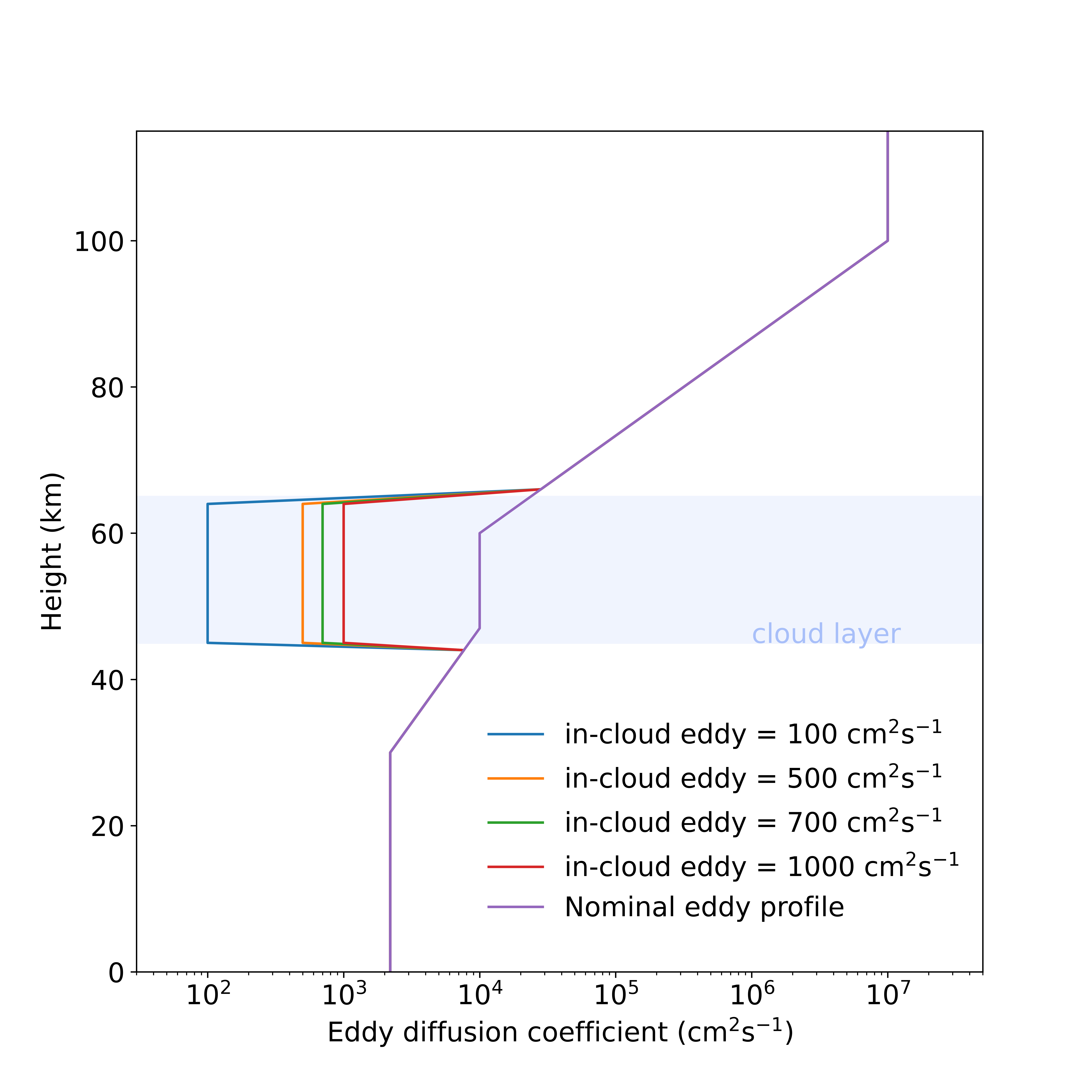}
 \caption{Vertical eddy diffusion profiles used as a function of altitude in Venus’s atmosphere. The nominal profile (turquoise) follows that of \citet{rimmer2021hydroxide} and \citet{2007Krasnopolsky}. Modified profiles vary the in-cloud eddy diffusion coefficient within the cloud layer (45–65~km altitude), adopting values of 100, 500, 700, and 1000~cm$^{2}$~s$^{-1}$ (orange, purple, and green respectively). Below and above the cloud layer, all profiles follow the same nominal structure.}
 \label{fig:eddy_profs}
\end{figure}

\begin{figure}[htbp]
	\includegraphics[width=\columnwidth]{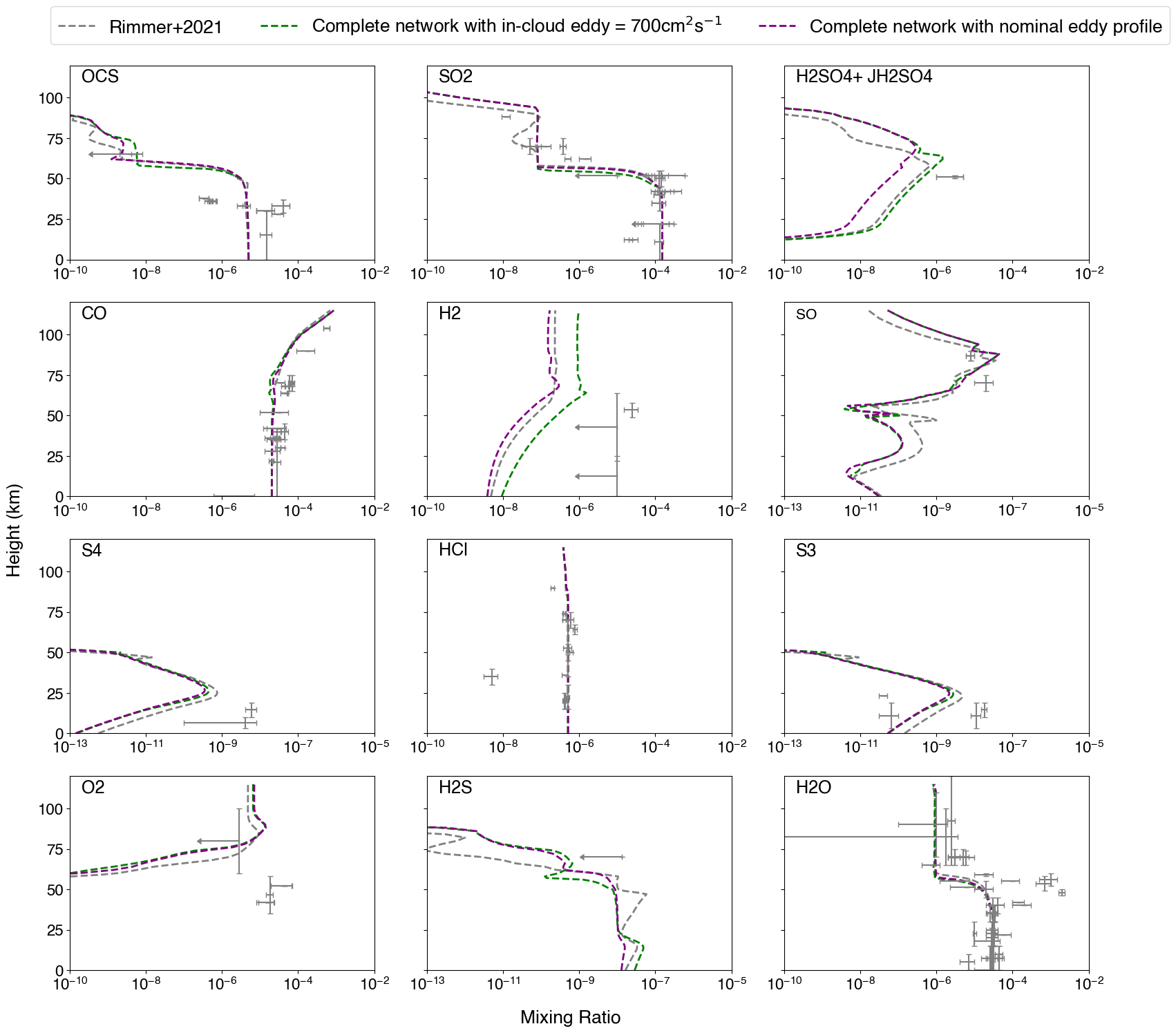}
 \caption{Predicted vertical profiles of \ce{SO2}, \ce{H2O}, \ce{CO}, \ce{OCS}, \ce{O2}, \ce{HCl}, \ce{H2SO4}, \ce{H2}, \ce{SO}, \ce{S3}, \ce{S4}, and \ce{H2S} from three atmospheric chemistry models. The \citet{rimmer2021hydroxide} model is shown with a grey dashed line; the complete chemical network (\citet{rimmer2021hydroxide} updated with some NO-reactions from \citet{Krasnopolsky2012}) with the nominal eddy profile is shown in purple dashed; and the complete network with a modified in-cloud (45--65\,km) eddy diffusion coefficient of 700~cm$^{2}$~s$^{-1}$ is shown in green dashed. Observational constraints and upper limits for each species, compiled in \citet{rimmer2021hydroxide}, are plotted in grey for comparison. The sulphuric acid profile includes both gaseous (\ce{H2SO4}) and condensed phases (\ce{JH2SO4}).
}
 \label{fig:comp_all_obs}
\end{figure}

\begin{figure}[htbp]
	\includegraphics[width=\columnwidth]{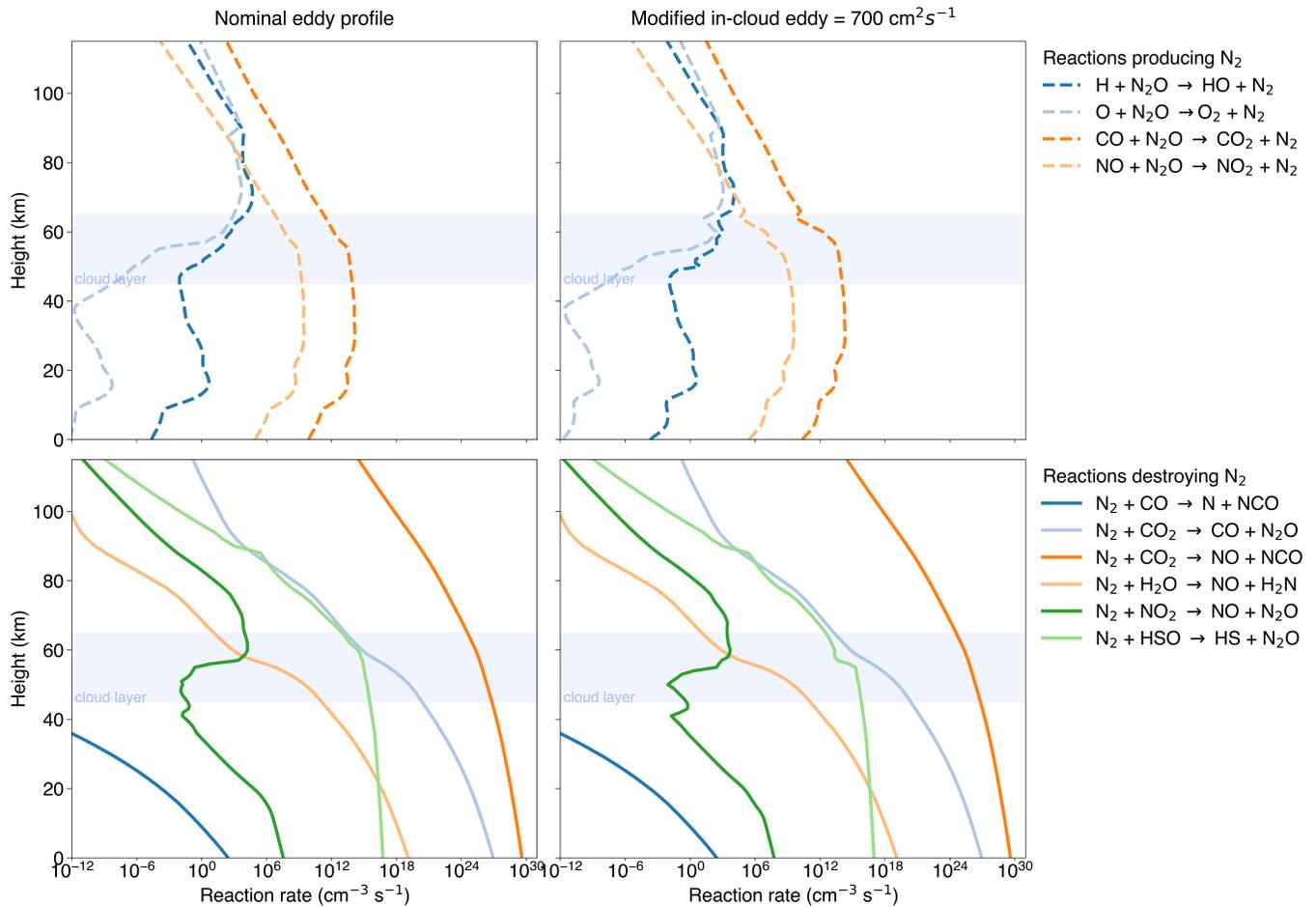}
 \caption{The rates of the dominant chemical reactions producing (top subplots, dashed lines) and destroying (bottom subplots, solid lines) N$_2$ as a function of altitude, for each eddy profile: the nominal (left subplots), and the modified eddy profile with in-cloud coefficients of 700\,cm$^{2}$~s$^{-1}$ that meets observational constraints (right subplots). Rates represent instantaneous values at the final time step for each altitude and do not capture the complete chemical history.}
 \label{fig:reactions_N2}
\end{figure}

\begin{figure}[htbp]
	\includegraphics[width=\columnwidth]{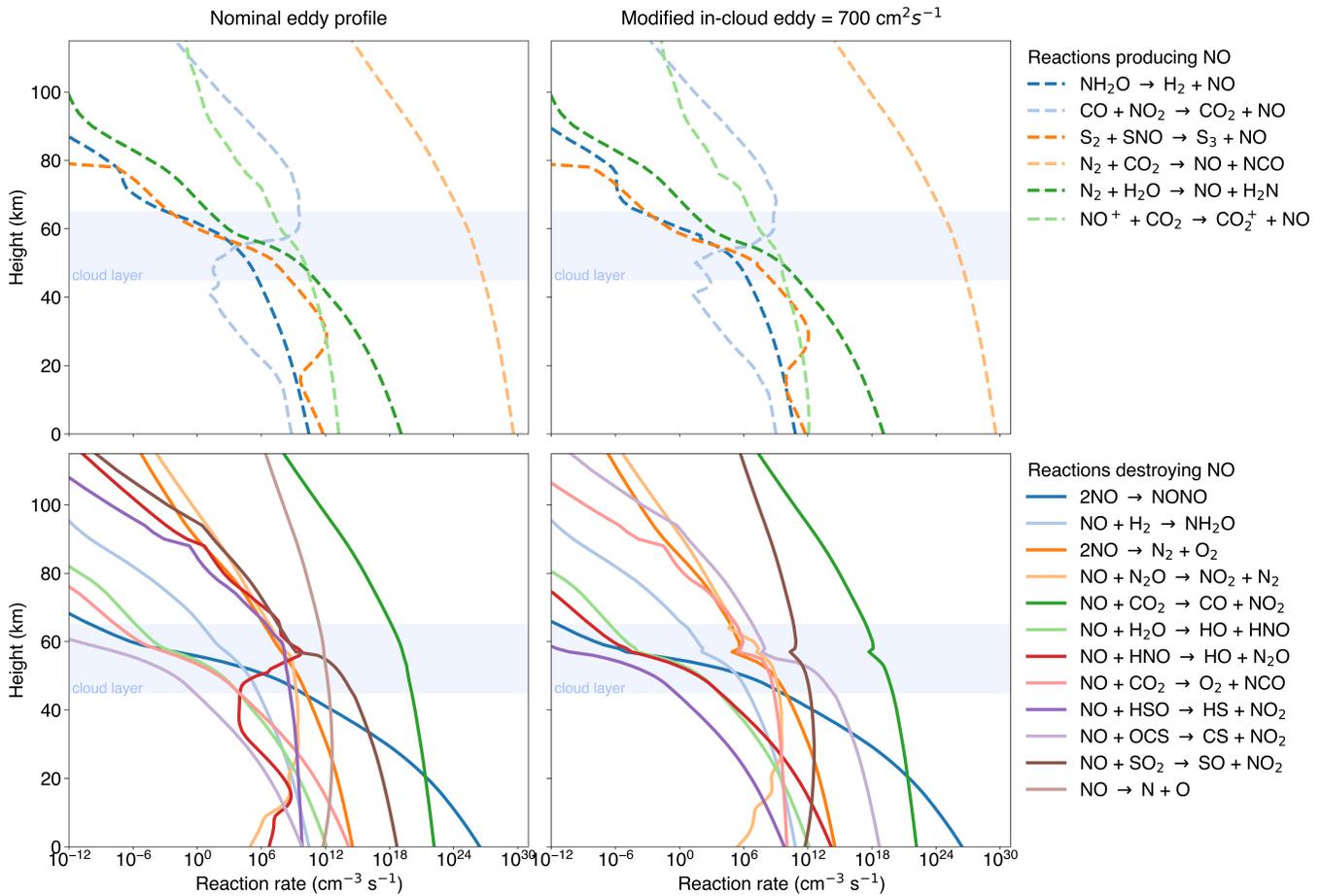}
 \caption{The rates of the dominant chemical reactions producing (top subplots, dashed lines) and destroying (bottom subplots, solid lines) NO as a function of altitude, for each eddy profile: the nominal (left subplots), and the modified eddy profile with in-cloud coefficients of 700\,cm$^{2}$~s$^{-1}$ that meets observational constraints (right subplots). Rates represent instantaneous values at the final time step for each altitude and do not capture the complete chemical history.}
 \label{fig:no_react}
\end{figure}

\begin{figure}[htbp]
	\includegraphics[width=\columnwidth]{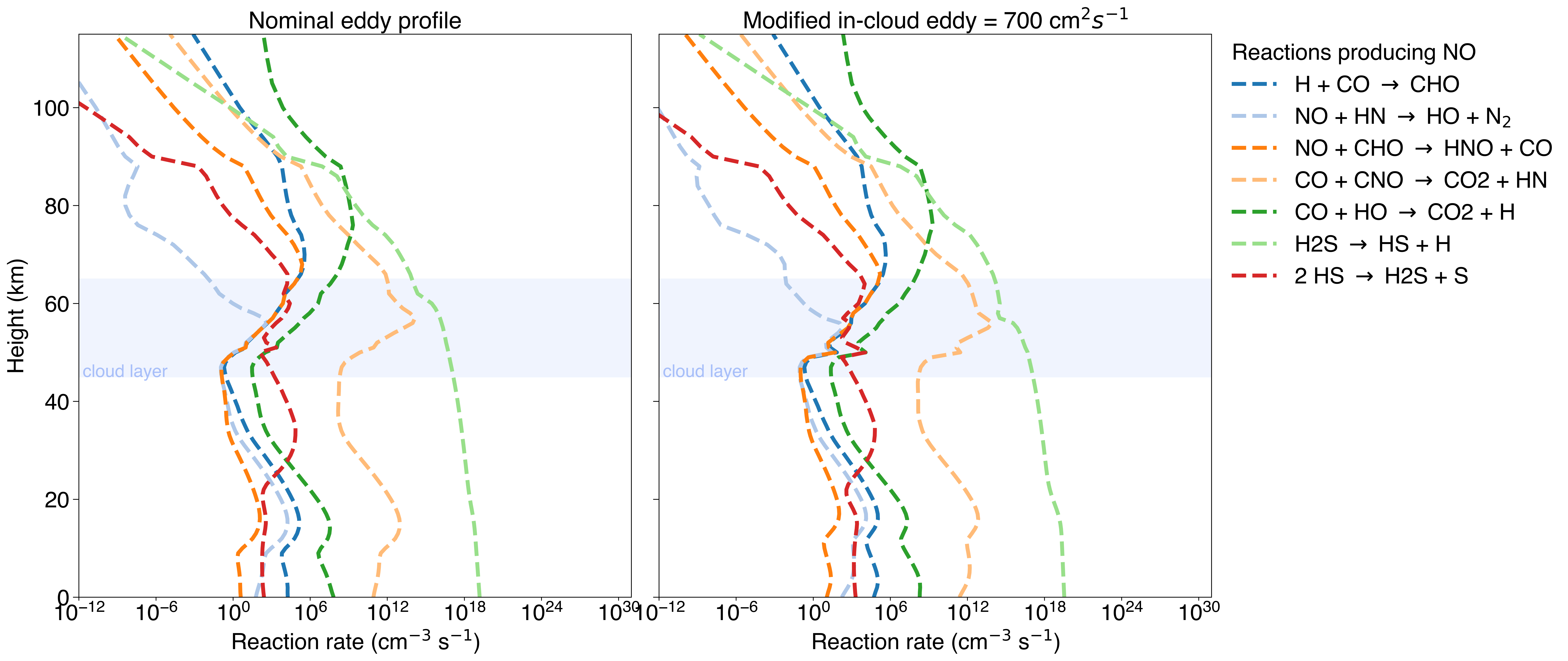}
 \caption{The rates of the chemical reactions  that make up the net depletion mechanism for NO, as a function of altitude, for each eddy profile: the nominal (left subplots), and the modified eddy profile with in-cloud coefficients of 700\,cm$^{2}$~s$^{-1}$ that meets observational constraints (right subplots). Rates represent instantaneous values at the final time step for each altitude and do not capture the complete chemical history.
}
 \label{fig:rate_limigint}
\end{figure}

\bibliography{references}{}
\bibliographystyle{aasjournalv7}


\end{document}